\documentclass[final,1p,times,sort&compress]{elsarticle}

\usepackage{amssymb}
\usepackage{amsmath}
\usepackage[colorlinks,linkcolor=blue,citecolor=blue,urlcolor=blue]{hyperref}
\hypersetup{pdfstartview={XYZ null null 1.15}}
\journal{Annals of Physics}

\renewcommand{\S}[1]{\sin(\theta_{#1})}
\newcommand{\Ssq}[1]{\sin^2(\theta_{#1})}
\newcommand{\Sh}[1]{\sin(\theta_{#1}/2)}
\newcommand{\Shsq}[1]{\sin^2(\theta_{#1}/2)}
\newcommand{\C}[1]{\cos(\theta_{#1})}
\newcommand{\Csq}[1]{\cos^2(\theta_{#1})}
\newcommand{\Ch}[1]{\cos(\theta_{#1}/2)}
\newcommand{\Chsq}[1]{\cos^2(\theta_{#1}/2)}
\newcommand{\R}[1]{\mathrm{e}^{i\eta_{#1}}}
\newcommand{\Rc}[1]{\mathrm{e}^{-i\eta_{#1}}}

\begin{document}

\begin{frontmatter}

\title{Inhomogeneous mean-field approach to collective excitations near the superfluid-Mott glass transition}

\author[Rolla,Regensburg]{Martin Puschmann}
\author[Rolla,SaoCarlos]{Jo\~ao C. Getelina}
\author[SaoCarlos]{Jos\'e A. Hoyos}
\author[Rolla]{Thomas Vojta}

\affiliation[Rolla]{organization={Department of Physics, Missouri University of Science and Technology},
            city={Rolla},
            state={Missouri},
            postcode={65409},
            country={USA}}

\affiliation[Regensburg]{organization={Institute of Theoretical Physics, University of Regensburg},
            postcode={D-93040},
            city={Regensburg},
            country={Germany}}

\affiliation[SaoCarlos]{organization={Instituto de F\'{i}sica de S\~ao Carlos, Universidade de S\~ao Paulo, C.P. 369},
            city={S\~ao Carlos},
            state={S\~ao Paulo},
            postcode={13560-970},
            country={Brazil}}

\begin{abstract}
We develop an inhomogeneous quantum mean-field approach to the behavior of collective excitations
across the superfluid-Mott glass quantum phase transition in two dimensions, complementing
recent quantum Monte Carlo simulations [Phys. Rev. Lett. {\bf 125}, 027002 (2020)].
In quadratic (Gaussian) approximation, the Goldstone (phase)
and Higgs (amplitude) modes completely decouple. Each is described by a disordered Bogoliubov
Hamiltonian which can be solved by an inhomogeneous multi-mode Bogoliubov transformation.
We find that the Higgs mode is spatially localized in both phases. The corresponding
scalar spectral function shows a broad peak that is noncritical in the sense that its peak
frequency does not soften but remains nonzero across the quantum phase transition.
In contrast, the lowest-energy Goldstone mode delocalizes in the superfluid phase, leading to
a zero-frequency spectral peak. We compare these findings to the results of the quantum Monte Carlo
simulations. We also relate them to general results on the localization of bosonic excitations,
and we discuss the limits and generality of our approach.
\end{abstract}

\begin{keyword}
quantum phase transition \sep disorder \sep collective excitation \sep superfluid \sep localization
\end{keyword}

\end{frontmatter}


\section{Introduction}
\label{sec:introduction}
%%%%%%%%%%%%%%%%%%%%%%%%%%%%%%%%%%%%%%%%%%%%%%%%%%%%%%%%%%%%%%%%%%%%%%%%%%%%%%%%%%%%%

Systems of disordered interacting bosons find diverse experimental applications
such as helium absorbed in porous media \cite{CHSTR83,Reppy84},
thin superconducting films \cite{HavilandLiuGoldman89,HebardPaalanen90}, Josephson
junction arrays \cite{ZFEM92,ZEGM96}, ultracold gases in optical
lattices \cite{WPMZCD09,KSMBE13,DTGRMGIM14}, and certain disordered magnetic materials
\cite{OosawaTanaka02,HZMR10,Yuetal12,Huevonenetal12,ZheludevRoscilde13}.
Their zero-temperature phase transitions between superfluid and insulating quantum
ground states are prototypical disordered quantum phase transitions.

The influence of impurities, defects, and other types of quenched disorder on
quantum phase transitions has been analyzed extensively since the 1990's. The resulting
body of work has established that quantum phase transitions are generally more strongly
affected by quenched disorder than their classical counterparts.
Unconventional phenomena such as infinite-randomness criticality  \cite{Fisher92,Fisher95},
smeared phase transitions \cite{Vojta03a,HoyosVojta08}, and quantum Griffiths singularities
\cite{ThillHuse95,RiegerYoung96} appear in a variety of systems. They can be
classified \cite{MMHF00,VojtaSchmalian05,VojtaHoyos14} based on the symmetries of the
order parameter and the defects, on the order parameter dynamics, and on the importance
of rare fluctuations (for reviews see, e.g., Refs.\ \cite{Vojta06,Vojta10,Vojta19}).

The majority of the past theoretical work on disordered quantum phase transitions was focused
on the behavior of thermodynamic quantities across the transition. The character and dynamics
of excitations has been explored less, despite the fact that they govern a variety of experiments
ranging from inelastic neutron scattering in magnetic materials to various electrical and
thermal transport measurements. Collective excitations due to the spontaneous breaking of
a continuous symmetry are particularly important. They include Goldstone modes
related to oscillations of the order parameter direction and amplitude (Higgs) modes
related to oscillations of the order parameter magnitude  (see, e.g., Refs.\
\cite{Burgess00,PekkerVarma15}).

Recently, Monte Carlo simulations were employed to investigate the behavior of the Higgs mode
across a paradigmatic disordered quantum phase transition, viz., the superfluid-Mott glass transition of
disordered interacting bosons  \cite{PuschmannCrewseHoyosVojta20,CrewseVojta21}. Even though the thermodynamic
critical behavior of this transition is known to be conventional \cite{ProkofevSvistunov04,Vojtaetal16},
the scalar susceptibility characterizing the Higgs mode was found to have unconventional
properties that violate naive scaling. Specifically, its spectral function $A^H(\omega)$
features a broad maximum whose peak frequency remains nonzero across the transition.
In the absence of disorder, in contrast, the scalar spectral function features a sharp Higgs
peak whose position, the Higgs energy (or mass), approaches zero at criticality, as expected
from scaling \cite{PodolskySachdev12,GazitPodolskyAuerbach13,Chenetal13}. What are the reasons
for the broad, noncritical scalar response (spectral function) at the superfluid-Mott glass
quantum phase transition?

To unravel and distinguish possible causes, such as enhanced damping of the Higgs mode or
spatial localization effects, a quantum mean-field theory was introduced in
Ref.\ \cite{PuschmannCrewseHoyosVojta20}. The purpose of the present paper is to fully
develop this mean-field theory and to explore its results for various disorder strengths and
distributions. Our approach generalizes the theories of Refs.\
\cite{AltmanAuerbach02,Pekkeretal12} to the disordered case and is also related to
the bond-operator method for disordered magnets \cite{VojtaM13}. Our theory captures
the spatially inhomogeneous character of the local order parameter in a disordered system.
Excitations are obtained from an expansion about the mean-field ground state.
More specifically, the Hamiltonian arising from expanding the Bose-Hubbard model to
quad\-ratic (Gaussian) order in the deviations from the mean-field ground contains two
completely decoupled sectors, representing the Goldstone and Higgs excitations, respectively.
Each sector takes the form of a disordered Bogoliubov Hamiltonian and can be solved
by an inhomogeneous multi-mode Bogoliubov transformation, giving us direct access to the
excitation energies and wave functions. At the Gaussian level, this approach produces
noninteracting bosonic excitations which implies that it captures localization effects
but not the broadening (damping) of the modes due to mode-mode interactions. This will us
allow us to separate possible causes of the unconventional scalar response.

The localization physics of noninteracting bosonic modes has been studied quite extensively
in the literature, with applications to phonons \cite{JohnSompolinskyStephen83},
electromagnetic waves (see, e.g., Ref.\ \cite{Sheng_book90}), and magnons
\cite{ChernyshevChenCastroNeto01}, among others. Generic aspects and symmetry considerations for
the localization of bosonic excitations were emphasized in Refs.\
\cite{GurarieChalker02,GurarieChalker03}. However, to the best of our knowledge,
not all of the questions raised above are answered in this body of literature,
in particular about the behavior of the response functions across the phase transition.
Moreover, it is interesting to ask whether the specific spatial correlations that
appear in the matrix elements of the fluctuation Hamiltonians affect the localization
properties.

Our main results can be summarized as follows. The mean-field ground state in
the superfluid phase features a spatially inhomogeneous local order parameter. The relative
spatial variations of the order parameter are small deep inside the superfluid phase,
but they grow as the transition is approached, reflecting the Griffiths region close to the
transition.
In the insulating (Mott) phase, the $U(1)$ order parameter symmetry is not broken, thus both
excitation sectors are degenerate and all modes are localized. In the superfluid phase,
Goldstone and Higgs excitations show qualitatively different behaviors. The lowest Goldstone
excitation (having zero excitation energy) is extended over the entire system, in agreement
with the general symmetry analysis \cite{GurarieChalker02,GurarieChalker03} and with
explicit results for Goldstone modes in a number of systems. In contrast, the lowest Higgs excitation
is localized in the superfluid phase. Importantly, the properties of
the Higgs spectral function $A^H(\omega)$ are analogous to the Monte Carlo results
\cite{PuschmannCrewseHoyosVojta20}, i.e., $A^H(\omega)$ displays a broad maximum whose
peak frequency remains nonzero across the transition. This suggests that localization effects
rather than enhanced damping are the (main) reason for the unconventional behavior of the
scalar response.

% These qualitative features are independent
% of the disorder strength and type (provided the disorder respects the particle-hole symmetry).

The rest of the paper is organized as follows. Section\ \ref{sec:BHM} defines the model.
The inhomogeneous quantum mean-field theory is developed in Sec.\ \ref{sec:mftheory}.
The effective Hamiltonians for the Goldstone and Higgs
excitations are derived in Sec.\ \ref{sec:fh}, and their diagonalization is discussed
in Sec.\ \ref{sec:bogoliubov}.
 Section \ref{sec:analysis} introduces the specific quantities and data analysis techniques
we employ to study the excitations. For comparison purposes, Sec.\ \ref{sec:clean} summarizes
the results for a clean square lattice Bose-Hubbard model.
Simulations for both the Goldstone and the Higgs mode are discussed in Sec.\ \ref{sec:simulations}.
We conclude in Sec.\ \ref{sec:conclusions}.

%%%%%%%%%%%%%%%%%%%%%%%%%%%%%%%%%%%%%%%%%%%%%%%%%%%%%%%%%%%%%%%%%%%%%%%%%%%%%%%%%%%%%
\section{Bose-Hubbard model}
\label{sec:BHM}
%%%%%%%%%%%%%%%%%%%%%%%%%%%%%%%%%%%%%%%%%%%%%%%%%%%%%%%%%%%%%%%%%%%%%%%%%%%%%%%%%%%%%

The Bose-Hubbard model (BHM) describes interacting bosons on a lattice. Its Hamiltonian reads
\begin{equation}
   H= -\frac 1 2 \sum_{ij} J_{ij} (a^\dagger_i a_j + \mathrm{h.c.}) + \frac{1}{2} \sum_i U_i (n_i-\bar{n})^2
   \label{eq:BHM}
\end{equation}
where $a_i^\dagger$ and $a_i$ are the boson creation and annihilation operator on lattice site $i$,
and $n_i=a_i^\dagger a_i$ is the boson number operator. $U_i$ is a repulsive on-site interaction, and
$J_{ij}$ denotes the hopping amplitude between sites $i$ and $j$. Later, we will focus on the square
 lattice with nearest-neighbor hopping only, but the theory can be developed without these restrictions.
$\bar{n}$ is the filling (or background ``charge''). A nonzero chemical potential
can be absorbed in a shift of $\bar{n}$. We are interested in the case of large integer $\bar{n}$
for which the Bose-Hubbard model is particle-hole symmetric.

In the clean case, consisting of an undiluted lattice with uniform $U_i \equiv U$
and translationally invariant $J_{ij} = f(\mathbf{r}_{ij})$
(where $f$ decays sufficiently rapidly with the distance $\mathbf{r}_{ij}$ between sites $i$ and $j$),
the qualitative behavior of the model is readily understood.
If the interactions dominate over the hopping terms, $J_{ij} \ll U$,
the ground state is a gapped, incompressible Mott insulator. In the opposite limit, $J_{ij} \gg U$,
the ground state is a superfluid. These two phases are separated by a continuous quantum phase transition
in the $(2+1)$ dimensional $XY$ universality class.

In the presence of quenched disorder, the superfluid and Mott insulator phases are always separated by an
insulating ``glass'' phase in which rare large regions of local superfluid order (superfluid ``puddles'')
coexist with the insulating bulk \cite{GiamarchiSchulz88,FisherFisher88,FWGF89,WeichmanMukhopadhyay07}.
This glass phase thus constitutes the quantum Griffiths phase of the superfluid-Mott insulator quantum phase
transition. The nature of the glassy intermediate phase depends on the symmetry properties of the disorder.
For generic disorder that locally breaks the particle-hole symmetry (realized, e.g., via a random potential
for the bosons), it is the so-called Bose glass, a compressible gapless insulator.
In the present paper, we focus on disorder that does not break the particle-hole symmetry.
In this case, the intermediate phase between superfluid and Mott insulator is not a Bose glass but rather
the incompressible gapless Mott glass \cite{GiamarchiLeDoussalOrignac01,WeichmanMukhopadhyay08}.

We employ two different types of the (particle-hole symmetry preserving) disorder.
The first disorder type is site dilution, i.e., we randomly remove a nonzero fraction $p$ of lattice
sites while the $U_i$ and $J_{ij}$ of the remaining sites stay translationally invariant.
This can be formally written as $U_i = \epsilon_i U$, $J_{ij} = \epsilon_i \epsilon_j f(\mathbf{r}_{ij})$
where the $\epsilon_i$ are independent random variables that take the values 0 (vacancy) with probability $p$
and 1 (occupied site) with probability $1-p$.

Within the second disorder type, the lattice is undiluted but the values of the onsite interactions $U_i$
are independent random variables drawn from a uniform  probability density of average $U$ and
relative width $r$,
\begin{equation}
P(U_i) = \left \{  \begin{array}{cl}
1/(r U) & \textrm{for} \quad (1-r/2)U < U_i < (1+r/2)U \\
0       & \textrm{otherwise}
\end{array} \right. ~
\label{eq:U_distrib}
\end{equation}
whereas the $J_{ij}$  remain translationally invariant.
The values of $r$ are restricted to $0\le r < 2$ to avoid finding sites where the
onsite interaction is zero or even attractive, $U_i < 0$. The strength of this ``random-$U$''
disorder can be tuned to zero continuously whereas the defects created by site dilution
always have a finite strength locally.

%%%%%%%%%%%%%%%%%%%%%%%%%%%%%%%%%%%%%%%%%%%%%%%%%%%%%%%%%%%%%%%%%%%%%%%%%%%%%%%%%%%%%
\section{Inhomogeneous mean-field theory}
\label{sec:mftheory}
%%%%%%%%%%%%%%%%%%%%%%%%%%%%%%%%%%%%%%%%%%%%%%%%%%%%%%%%%%%%%%%%%%%%%%%%%%%%%%%%%%%%%

Our approach follows Ref.\ \cite{AltmanAuerbach02,Pekkeretal12} and generalizes the theory to the disordered case.
 It is also related to the bond-operator method for disordered magnets \cite{VojtaM13}.
In the Mott-insulating phase, the local boson numbers $n_i$ fluctuate only weakly about their preferred
value $\bar{n}$. The problem is thus well approximated by truncating the local Hilbert space associated with
site $i$ to the space spanned by the boson number eigenstates $|\bar{n}-1\rangle_i$,
$|\bar{n}\rangle_i$, and $|\bar{n}+1\rangle_i$. This truncation
qualitatively captures the essential physics even in the superfluid phase.
The truncated local Hilbert space at site $i$ can be constructed employing three commuting
$t_{i,\alpha}$ bosons ($\alpha=-1,0,1$) forming the states $|\bar{n}+\alpha\rangle_i=t_{i,\alpha}^\dagger|\textrm{vac}\rangle$
out of the fictitious vacuum $|\textrm{vac}\rangle$. They fulfill the constraint $\sum_\alpha t_{i,\alpha}^\dagger t_{i,\alpha} =1$.
The original $a_i^\dagger$ boson creation operators of the Bose-Hubbard model (\ref{eq:BHM}) can be represented
as $a_i^\dagger= \sqrt{\bar{n}} t_{i,0}^\dagger t_{i,-1} + \sqrt{\bar{n}+1} t_{i,1}^\dagger t_{i,0}$.

In the limit of large filling, $\bar{n}\gg 1$, the Bose-Hubbard model takes the simple pseudo-spin-one form
\begin{equation}
	H_S = -\frac 1 4 \sum_{ij} \tilde J_{ij} (S^+_i S^-_j + \mathrm{h.c.}) + \frac 1 2 \sum_i U_i (S^z_i)^2~,
\label{eq:H_spin}
\end{equation}
with $\tilde J_{ij}=\bar n J_{ij}$. The spin operators are given by
\begin{align}
	S_i^+ &= \sqrt{2} \left(t_{i,1}^\dagger t_{i,0} + t_{i,0}^\dagger t_{i,-1}\right)=(S_i^-)^\dagger~, \label{eq:S+}\\
	S_i^z &= t_{i,1}^\dagger t_{i,1} - t_{i,-1}^\dagger t_{i,-1}~, \label{eq:Sz}
\end{align}
and the eigenstates of the $S_i^z$ operator, $|+\rangle_i$, $|0\rangle_i $, and $|-\rangle_i$ correspond
to the original boson number eigenstates $|\bar{n}+1\rangle_i$, $|\bar{n}\rangle_i$, and $|\bar{n}-1\rangle_i$,
respectively.

The inhomogeneous mean-field theory is based on a product ansatz for the ground state wave function
\begin{equation}
|\Phi_0\rangle=\prod_{j}|\phi_{0}\rangle_j
              =\prod_{j}\Big[\Ch{j}|0\rangle_j +\Sh{j}\left(\R{j}|+\rangle_j+\Rc{j}|-\rangle_j\right)/\sqrt{2}\Big]
\label{eq:gs_wavefunction}
\end{equation}
where $\theta_j$ and $\eta_j$ are local variational parameters ($0\leq\theta_j\leq\pi/2$, $0\leq\eta_j <2\pi$).
For $\theta_j=0$, the wave function
describes the Mott-insulating phase because the local boson number is fixed at $n_i = \bar n$, without any fluctuations.
For $\theta_j>0$, the wave function features a nonzero local superfluid order parameter
$\langle a_j^\dagger\rangle = \sqrt{\bar n}\, \psi_j$ with  $\psi_j = \langle S_j^+\rangle=\Rc{j}\S{j}$.
Thus, the mixing angle $\theta_j$ parameterizes the local order parameter amplitude and $\eta_j$ its phase.

Evaluating the expectation value of $H_S$ in the product state (\ref{eq:gs_wavefunction}) yields the variational
ground state energy
\begin{eqnarray}
	E_{0}=\langle \Phi_0 | H_s | \Phi_0 \rangle = \frac 1 2 \sum_{i} U_i \Shsq{i} - \frac 1 2 \sum_{ij} \tilde J_{ij} \S{i}\S{j} \cos(\eta_i -\eta_j)~. \label{eq:thermodnamic_groundstate}
\end{eqnarray}
It is minimized by mixing angles fulfilling the coupled mean-field equations
\begin{equation}
	4\C{i}\sum_{j} \tilde J_{ij} \S{j} = U_i \S{i}
\label{eq:sf_groundstate_condition}
\end{equation}
while the phases are uniform, $\eta_i \equiv \eta= \textrm{const}$. In the following,
we set these phases to zero without loss of generality.
The trivial Mott-insulating state, $\S{i} = 0$,
is always a solution of the mean-field equations (\ref{eq:sf_groundstate_condition}).
For sufficiently large $\tilde J_{ij}$ or small $U_i$, a nontrivial superfluid
solution of (\ref{eq:sf_groundstate_condition}) may appear.

In the presence of disorder, the system of equations (\ref{eq:sf_groundstate_condition}) usually
needs to be solved numerically. In the clean case, in contrast, the system
(\ref{eq:sf_groundstate_condition}) reduces to a single equation which can be solved analytically.
An example will be presented in Sec.\ \ref{sec:clean}.

%%%%%%%%%%%%%%%%%%%%%%%%%%%%%%%%%%%%%%%%%%%%%%%%%%%%%%%%%%%%%%%%%%%%%%%%%%%%%%%%%%%%%
\section{Fluctuation Hamiltonians}
\label{sec:fh}
%%%%%%%%%%%%%%%%%%%%%%%%%%%%%%%%%%%%%%%%%%%%%%%%%%%%%%%%%%%%%%%%%%%%%%%%%%%%%%%%%%%%%

Having obtained the mean-field ground state, we now consider excitations on top of it. The goal is to derive
effective Hamiltonians for these excitations. As a first step, we introduce a new set of local boson operators
via the unitary transformation
\begin{eqnarray}
		\begin{pmatrix}
		b_{0,j}^\dagger\\
		b_{G,j}^\dagger\\
		b_{H,j}^\dagger
		\end{pmatrix}
        =
        		\begin{pmatrix}
		\frac{1}{\sqrt{2}}\Sh{j} & \Ch{j} & \frac{1}{\sqrt{2}}\Sh{j} \\
		\frac{i}{\sqrt{2}} & 0 & -\frac{i}{\sqrt{2}}  \\
		\frac{1}{\sqrt{2}}\Ch{j} &-\Sh{j} & \frac{1}{\sqrt{2}}\Ch{j}
		\end{pmatrix}
        \begin{pmatrix}
		t_{+,j}^\dagger\\
		t_{0,j}^\dagger\\
		t_{-,j}^\dagger
		\end{pmatrix},
\label{eq:tb-trafo}
\end{eqnarray}
such that $b_{0,j}^\dagger$ creates the local mean-field ground state out of the fictitious vacuum. $b_{G,j}^\dagger$ and $b_{H,j}^\dagger$ create states orthogonal to $b_{0,j}^\dagger |\textrm{vac}\rangle$ and describe, respectively, a change in the local
order parameter phase $\eta_j$ and a change in the local order parameter amplitude $\S{j}$. These operators
fulfill the constraint
\begin{equation}
b_{0,j}^\dagger b_{0,j} + b_{G,j}^\dagger b_{G,j} + b_{H,j}^\dagger b_{H,j} =1~.
\label{eq:b_constraint}
\end{equation}
To express the Bose-Hubbard Hamiltonian in terms $b$ bosons, we first rewrite the spin-one operators
\begin{align}
(S^z_j)^2=& \Shsq{j} n_{0,j} +n_{G,j}+\Chsq{j} n_{H,j} +\Sh{j}\Ch{j} (b_{0,j}^\dagger b_{H,j} + b_{H,j}^\dagger b_{0,j} ),\\
S^+_j=& \S{j} (n_{0,j}-n_{H,j})  + \C{j} ( b_{0,j}^\dagger b_{H,j} + b_{H,j}^\dagger b_{0,j}) \nonumber \\
      & \quad - i\Ch{j} (b_{0,j}^\dagger b_{G,j} + b_{G,j}^\dagger b_{0,j}) + i\Sh{j} (b_{G,j}^\dagger b_{H,j} + b_{H,j}^\dagger b_{G,j} ) .
\end{align}
We eliminate (``fully condense'') by means of the constraint (\ref{eq:b_constraint})
the operators $b_{0,j}^\dagger$ and $b_{0,j}$ that correspond to the mean-field ground state.
This is analogous to a Holstein-Primakoff formalism with
$n_{0,j}\gg n_{G,j}$ and $n_{0,j}\gg n_{H,j}$.
Expanding to quadratic order in $b_{G,j}$ and $b_{H,j}$, the spin-one operators can be written as
\begin{eqnarray}
(S^z_j)^2 &=& \Shsq{j} + \Chsq{j} n_{G,j}+\C{j} n_{H,j} +\Sh{j}\Ch{j} (b_{H,j}+b_{H,j}^\dagger)  ~,~\\
S^+_j &=&  \S{j} (1-n_{G,j}-2 n_{H,j})  + \C{j} (b_{H,j}+b_{H,j}^\dagger) - i \Ch{j}(b_{G,j}+b_{G,j}^\dagger)~.
\end{eqnarray}

Inserting these representations into the pseudo-spin Hamiltonian (\ref{eq:H_spin}), and using the mean-field equations
(\ref{eq:sf_groundstate_condition}) to cancel out terms linear in $b_{G,j}$ and $b_{H,j}$, the Hamiltonian becomes,
to quadratic order in $b_{G,j}$ and $b_{H,j}$,
the sum of the mean-field ground state energy and two fluctuation contributions,
$H=E_0+\mathcal{H}_{G}+\mathcal{H}_{H}$. The fluctuation Hamiltonians read
\begin{eqnarray}
\mathcal{H}_{G} &=& - \frac 1 2 \sum_{ij} \tilde J_{ij} \Ch{i} \Ch{j} (b_{G,i}+b_{G,i}^\dagger) (b_{G,j}+b_{G,j}^\dagger)
         + \sum_{i}\varpi_{G,i} n_{G,i} ~, \label{eq:H_G} \\
\mathcal{H}_{H} &=& - \frac 1 2 \sum_{ij} \tilde J_{ij} \C{i} \C{j} (b_{H,i}+b_{H,i}^\dagger) (b_{H,j}+b_{H,j}^\dagger)
         + \sum_{i}\varpi_{H,i} n_{H,i}. \label{eq:H_H}
\end{eqnarray}
The  Goldstone and Higgs modes thus completely decouple to quadratic order in $b_{G,j}$ and $b_{H,j}$.
$\mathcal{H}_{G}$ and $\mathcal{H}_{H}$ each take the form of a disordered Bogoliubov Hamiltonian
describing a set of coupled harmonic oscillators with the local frequencies
$\varpi_{G,i}= (U_i/2)\Chsq{i}+\zeta_i$ and $\varpi_{H,i}= (U_i/2)\C{i}+2\zeta_i$ where  $\zeta_i=\S{i}\sum_{j}\tilde J_{ij}\S{j}$.

In the Mott-insulating phase, the local mixing angles $\theta_i$ all vanish. Thus
$\mathcal{H}_{G}$ and $\mathcal{H}_{H}$ are identical, in agreement with the fact that
the two excitation modes are degenerate if the $U(1)$ order parameter symmetry is not broken.

%%%%%%%%%%%%%%%%%%%%%%%%%%%%%%%%%%%%%%%%%%%%%%%%%%%%%%%%%%%%%%%%%%%%%%%%%%%%%%%%%%%%%
\section{Bogoliubov transformation}
\label{sec:bogoliubov}
%%%%%%%%%%%%%%%%%%%%%%%%%%%%%%%%%%%%%%%%%%%%%%%%%%%%%%%%%%%%%%%%%%%%%%%%%%%%%%%%%%%%%

Due to the presence of anomalous $b^\dagger b^\dagger$ and $bb$ terms, the fluctuation Hamiltonians
$\mathcal{H}_{G}$ and $\mathcal{H}_{H}$ have to be solved by means of a multi-modal
Bogoliubov transformation ($\alpha=G,H$)
\begin{eqnarray}
b_{\alpha j} = \sum_n ( u_{\alpha jn} d_{\alpha n} + v_{\alpha jn}^\ast d_{\alpha n}^\dagger )
\label{eq:bogoliubov}
\end{eqnarray}
where the $d$ (Bogoliubov) bosons correspond to the excitation eigenstates of our system.

The transformation coefficients $u$ and $v$ can be found efficiently by interpreting each of the
fluctuation Hamiltonians as a system of coupled harmonic oscillators and switching to a
first-quantization framework. The local oscillator energies in eqs.\ (\ref{eq:H_G})
and (\ref{eq:H_H}) can be written as
$\varpi_{\alpha,i} (n_{\alpha,i} + 1/2) = p_{\alpha,i}^2/2 + \varpi_{\alpha,i} x_{\alpha,i}^2/2$
where $p_{\alpha,i}$ and $x_{\alpha,i}$ represent the momentum and position operators
of the fictitious oscillator at the site $i$. (The mass can be set to unity without loss of generality.)
The terms $(b_{\alpha,i}+b_{\alpha,i}^\dagger)$ that appear in the intersite couplings in
eqs.\ (\ref{eq:H_G}) and (\ref{eq:H_H}) can be expressed as
$(b_{\alpha,i}+b_{\alpha,i}^\dagger) = \sqrt{2\varpi_{\alpha,i}} x$.
In first quantization, the fluctuation Hamiltonians $\mathcal{H}_{G}$ and $\mathcal{H}_{H}$
thus take the form
\begin{eqnarray}
	\mathcal{H}_\alpha &=&  \sum_{i} \frac{p_{\alpha,i}^2}{2} + \sum_{ij} \frac{x_{\alpha,i} {X}_{\alpha,ij} x_{\alpha,j}}{2}~.
    \label{eq:H_first}
\end{eqnarray}
The coupling matrices are given by
\begin{eqnarray}
	X_{G,ij} &=& \varpi^2_{G,i} \delta_{ij} -2 \Ch{i} \Ch{j} \tilde J_{ij}  \sqrt{\varpi_{G,i}\varpi_{G,j}}  \label{eq:X_G}\\
	X_{H,ij} &=& \varpi^2_{H,i} \delta_{ij} - 2 \C{i} \C{j} \tilde J_{ij}  \sqrt{\varpi_{H,i}\varpi_{H,j}} \label{eq:X_H}
\end{eqnarray}
for the Goldstone and Higgs modes, respectively. The Hamiltonians (\ref{eq:H_first}) are now
easily solved by diagonalizing the coupling matrices $\mathbf{X}_{\alpha}$ which are real symmetric
$N\times N$ matrices in a system of $N$ lattice sites,
\begin{eqnarray}
	\sum_k X_{\alpha, jk} \mathcal{V}_{\alpha, kn} = \nu_{\alpha, n}^2 \mathcal{V}_{\alpha, jn}
  \label{eq:EV_problem_first}
\end{eqnarray}
where $\nu_{\alpha, n}$ is the $n$-th nonnegative excitation eigenfrequency (energy), and
the $n$-th column of the matrix $\mathcal{V}_{\alpha, jn}$ contains the $n$-th eigenvector. Going back to second
quantization in the eigenbasis of $\mathbf{X}_{\alpha}$ yields
\begin{equation}
\mathcal{H}_G= \sum_n \nu_{G,n} d_{G,n}^\dagger d_{G,n} ~, \qquad  \mathcal{H}_H= \sum_n \nu_{H,n} d_{H,n}^\dagger d_{H,n}
\end{equation}
with the $d$  bosons given by
\begin{eqnarray}
	d_{\alpha,n} &=& \sqrt{\frac{\nu_{\alpha,n}}{2}}\left(\tilde{x}_{\alpha,n}+\frac{i}{\nu_{\alpha,n}}\tilde{p}_{\alpha,n}\right)
        =\sqrt{\frac{\nu_{\alpha,n}}{2}}\sum_{j}\mathcal{V}_{\alpha,jn}\left({x}_{\alpha,j}+\frac{i}{\nu_{\alpha,n}}{p}_{\alpha,j}\right) \nonumber\\
	&=& \frac{1}{2} \sum_{j}\mathcal{V}_{\alpha,jn}\left[\left(\sqrt{\frac{\nu_{\alpha,n}}{\varpi_{\alpha,j}}}
              -\sqrt{\frac{\varpi_{\alpha,j}}{\nu_{\alpha,n}}}\right)b_{\alpha,j}^\dagger
              +\left(\sqrt{\frac{\nu_{\alpha,n}}{\varpi_{\alpha,j}}}+\sqrt{\frac{\varpi_{\alpha,j}}{\nu_{\alpha,n}}}\right)b_{\alpha,j}
              \right] ~. \label{eq:dbosons_definition}
\end{eqnarray}
Here, $\tilde{x}_{\alpha,n}$ and $\tilde{p}_{\alpha,n}$ are the position and momentum operators in the eigenbasis of $\mathbf{X}_{\alpha}$.
Using the definitions $R_{\alpha,jn}=\mathcal{V}_{\alpha,jn}\sqrt{\nu_{\alpha,n}/\varpi_{\alpha,j}}$ and $L_{\alpha,jn}=\mathcal{V}_{\alpha,jn}\sqrt{\varpi_{\alpha,j}/\nu_{\alpha,n}}$,  the multi-modal Bogoliubov transformation
can be written in the compact form
\begin{equation}
	\begin{pmatrix}
	\mathbf{d}_\alpha \\
	\mathbf{d}_\alpha^\dagger
	\end{pmatrix}
	=\frac{1}{2}
	\begin{pmatrix}
	\mathbf{R}_\alpha^\mathrm{T}+ \mathbf{L}_\alpha^\mathrm{T} & \mathbf{R}_\alpha^\mathrm{T} - \mathbf{L}_\alpha^\mathrm{T} \\
	\mathbf{R}_\alpha^\mathrm{T}- \mathbf{L}_\alpha^\mathrm{T} & \mathbf{R}_\alpha^\mathrm{T} + \mathbf{L}_\alpha^\mathrm{T}
	\end{pmatrix}\begin{pmatrix}
	\mathbf{b}_\alpha \\
	\mathbf{b}_\alpha^\dagger
	\end{pmatrix} ~, \qquad\qquad
	\begin{pmatrix}
	\mathbf{b}_\alpha \\
    \mathbf{b}_\alpha^\dagger
	\end{pmatrix}
	=\frac{1}{2}
	\begin{pmatrix}
	\mathbf{L}_\alpha + \mathbf{R}_\alpha & \mathbf{L}_\alpha - \mathbf{R}_\alpha \\
	\mathbf{L}_\alpha - \mathbf{R}_\alpha & \mathbf{L}_\alpha + \mathbf{R}_\alpha
	\end{pmatrix}
	\begin{pmatrix}
    \mathbf	{d}_\alpha \\
	\mathbf{d}_\alpha^\dagger
	\end{pmatrix}\;,
\end{equation}
implying that the transformation coefficients $u$ and $v$ in eq.\ (\ref{eq:bogoliubov})
are given by $u_{\alpha,jn}= (L_{\alpha,jn} + R_{\alpha,jn})/2$ and
$v_{\alpha,jn}= (L_{\alpha,jn} - R_{\alpha,jn})/2$.

The ground states $|GS\rangle$ of the fluctuation Hamiltonians $\mathcal{H}_G$ and $\mathcal{H}_H$ are defined
by the absence of $d$ bosons, i.e., they correspond to $d$ vacuums. Expressing the $b$ operators in terms
of the $d$ operators shows that the ground states are \emph{not} $b$ vacuums. Instead, the $b$ number operator has the
ground state expectation value
\begin{equation}
\langle GS | b_{\alpha,j}^\dagger b_{\alpha,j} | GS\rangle = \frac 1 4 \sum_n (L_{\alpha,jn} - R_{\alpha,jn})^2~.
\label{eq:b_fluc}
\end{equation}

As discussed in Ref.\ \cite{GurarieChalker03}, the fluctuation Hamiltonians $\mathcal{H}_\alpha$ in
first-quantized form (\ref{eq:H_first}) can be mapped onto the standard form of a chiral symmetry class
Hamiltonian
\begin{equation}
\begin{pmatrix} 0 & Q \\ Q^\dagger & 0 \end{pmatrix} 
\end{equation}
where $Q$ is a real matrix.
The fluctuation Hamiltonians thus fulfill the symmetries of the chiral orthogonal class
(BDI in the Altland-Zirnbauer classification) \cite{EversMirlin08}. This also implies that
$v_{\alpha}=0$ is a special reference energy. Note, however, that the correlations
induced by the presence of the local order parameters in the matrix elements may potentially give rise
to nontrivial behavior beyond that of a chiral random matrix ensemble.

%%%%%%%%%%%%%%%%%%%%%%%%%%%%%%%%%%%%%%%%%%%%%%%%%%%%%%%%%%%%%%%%%%%%%%%%%%%%%%%%%%%%%
\section{Observables and data analysis}
\label{sec:analysis}
%%%%%%%%%%%%%%%%%%%%%%%%%%%%%%%%%%%%%%%%%%%%%%%%%%%%%%%%%%%%%%%%%%%%%%%%%%%%%%%%%%%%%

In order to study the character and dynamics of the excitations, we analyze the eigenfunctions
$\mathcal{V}_{\alpha, jk}$, and we compute a number of additional observables.

\subsection{Dynamic susceptibilities}
\label{subsec:dyn_susc}

We are interested in the longitudinal and transversal order parameter susceptibilities
as well as in the scalar susceptibility (the susceptibility of the order parameter
magnitude). In terms of the pseudo-spin-1 operators introduced in eqs.\
(\ref{eq:S+}) and (\ref{eq:Sz}), the longitudinal local order parameter component
is given by $S_j^x = (S_j^+ + S_j^-)/2$ (because we have fixed the phase $\eta_j$ of the
order parameter at zero in the mean-field solution). The transversal component
is given by $S_j^y = (S_j^+ - S_j^-)/(2i)$, and the (squared) order parameter amplitude
is associated with $ (S_j^x)^2 + (S_j^y)^2 = 2- (S_j^z)^2$.
For each of these operators, we define the retarded Green function. The longitudinal
susceptibility reads
\begin{equation}
G^\parallel_{jk} = -i \Theta(t) \langle GS |\, [S_j^x(t),S_k^x(0)] \, | GS \rangle~.
\label{eq:Green_long_definition}
\end{equation}
The transversal and scalar susceptibilities, $G^\perp_{jk}$ and $G^S_{jk}$,
are defined analogously. Expanding the operators to quadratic order in the
$b$ bosons, the longitudinal, transversal and scalar susceptibilities
\begin{eqnarray}
	G^\parallel_{jk}(t) &=&\C{i}\C{j}\, G^H_{jk}(t)\;, \label{eq:Green_long}\\
	G^\perp_{jk}(t) &=&\Ch{i}\Ch{j}\, G^{G}_{jk}(t)\;, \label{eq:Green_perp}\\
	G^\mathrm{S}_{jk}(t) &=&\frac{1}{4} \S{i} \S{j}\, G^H_{jk}(t) \label{eq:Green_scal}
\end{eqnarray}
can be decomposed into functions of the local order parameters (mixing angles $\theta_j$) and the Green functions
of the elementary excitations in the Higgs and Goldstone channels which are defined as
\begin{eqnarray}
	G^{G}_{jk}(t)=-i \Theta(t) \langle GS |\, [b_{G,j}(t)+b^\dagger_{G,j}(t), b_{G,k}(0)+b^\dagger_{G,k}(0)] \, | GS \rangle~,
    \label{eq:Green_G} \\
	G^{H}_{jk}(t)=-i \Theta(t) \langle GS |\, [b_{H,j}(t)+b^\dagger_{H,j}(t), b_{H,k}(0)+b^\dagger_{H,k}(0)] \, | GS \rangle~.
    \label{eq:Green_H}
\end{eqnarray}
To evaluate the ground state expectation values, we transform the $b$ bosons to the Bogoliubov
$d$ bosons by means of eq.\ (\ref{eq:dbosons_definition}), yielding
$b_{\alpha,j}+b^\dagger_{\alpha,j} = \sum_n L_{\alpha,jn} (d_{\alpha,n}+d^\dagger_{\alpha,n} )$.
After Fourier-transforming $G^{G}_{jk}(t)$ and $G^{H}_{jk}(t)$ w.r.t. time, the spectral functions $A^{G}_{jk}(\omega)$ and
$A^{H}_{jk}(\omega)$ are obtained as
\begin{eqnarray}
A^\alpha_{jk}(\omega)= -\frac 1 \pi \textrm{Im}  G^\alpha_{jk}(\omega)
               &=& \sum_n L_{\alpha,jn} L_{\alpha,kn}  [\delta(\omega - \nu_{\alpha,n}) - \delta(\omega + \nu_{\alpha,n}) ] \\
               &=& \sum_n \frac{\sqrt{\varpi_{\alpha,j}\varpi_{\alpha,k}}}{\nu_{\alpha,n}} \mathcal{V}_{\alpha,jn} \mathcal{V}_{\alpha,kn}
                               [\delta(\omega - \nu_{\alpha,n}) - \delta(\omega + \nu_{\alpha,n}) ] ~.
\label{eq:Ajk}
\end{eqnarray}
In the presence of disorder, the spectral functions $A^\alpha_{jk}(\omega)$ are not translationally invariant in space.
Translational invariance is restored after an ensemble average over disorder configurations. We can then perform a spatial
Fourier transformation to wave vector $\mathbf{q}$,
\begin{equation}
A^\alpha_{\mathbf{q}}(\omega)= \frac 1 N \sum_{j,k} e^{i (\mathbf{r}_j - \mathbf{r}_k) \cdot \mathbf{q}} \langle A^\alpha_{jk}(\omega) \rangle
\label{eq:A_q}
\end{equation}
where $\mathbf{r}_j$ is the position vector of site $j$, and $\langle \ldots \rangle$ denotes the disorder average.

Let us briefly discuss some qualitative features of the longitudinal, transversal,
and scalar susceptibilities. In the Mott-insulating phase, the Goldstone and Higgs
Green functions $G^G_{jk}(t)$ and $G^H_{jk}(t)$ are identical because the corresponding fluctuation
Hamiltonians agree with each other, as discussed at the end of Sec. \ref{sec:fh}.
As the local mixing angles $\theta_i$ all vanish in the Mott insulator,
eqs.\ (\ref{eq:Green_long}) and (\ref{eq:Green_perp}) imply
that the longitudinal and transversal susceptibilities coincide,
$G^\parallel_{jk}(t) = G^\perp_{jk}(t)$, in agreement with the fact that the
order parameter symmetry is not broken in the insulator phase.
In contrast, the scalar susceptibility $G^S_{jk}(t)$ vanishes in the Mott insulating
phase. (This is an artefact of the Gaussian approximation, as was already noted in
Ref.\ \cite{Pekkeretal12} for the clean case.) In the superfluid phase,
these susceptibilities all differ from each other.

\subsection{Localization properties: multifractal analysis}
\label{subsec:multifractal}

Because the Goldstone and Higgs excitations are the eigenstates of the spatially disordered
fluctuation Hamiltonians $\mathcal{H}_G$ and $\mathcal{H}_H$, they
are expected to feature nontrivial localization properties analogous to those observed
for other noninteracting bosonic excitations in disordered systems
\cite{JohnSompolinskyStephen83,Sheng_book90,ChernyshevChenCastroNeto01,GurarieChalker02,GurarieChalker03}.
 We analyze these properties
by means of two methods, (i) a multifractal analysis of the eigenstates
and (ii) the scaling of the Lyapunov exponent in a quasi-one-dimensional geometry
found via the recursive Green function method.

Within the multifractal analysis
\cite{CastellaniPeliti86,SchreiberGrussbach91,VasquezRodriguezRoemer08,RodriguezVasquezRoemer08},
we partition the system defined on a square lattice of $L^2$ sites
into square boxes of linear size $l$.  For each eigenstate $n$ of the fluctuation Hamiltonians
$\mathcal{H}_\alpha$ (with $\alpha=G,H$), we define a measure $\mu_j(n,l,L)$
characterizing the probability of the eigenstate in a box of size $l$ with lower left corner
at site $j$. It reads
\begin{equation}
\mu_j(n,l,L) = \sum_{k \in \textrm{box}} |\mathcal{V}_{\alpha, kn}|^2 =
     \sum_{k \in \textrm{box}} (|u_{\alpha, kn}|^2 - |v_{\alpha, kn}|^2)
\end{equation}
where the sum runs over all lattice sites $k$ in the box. We then construct the $q$-th moment of these
box probabilities,
\begin{equation}
P_q (n,l,L) = \frac 1 {l^2} \sum_j \mu_j^q(n,l,L)~.
\end{equation}
The sum runs over all sites $j$, i.e., it considers all possible (overlapping) boxes of size $l$.
The prefactor $1/l^2$ guaranties the proper normalization of the probability, $P_1(n,l,L) = 1$.
Note that $P_2(n,1,L)$ corresponds to the inverse participation number of state $n$.

For multifractal wave functions\footnote{Multifractality is usually a property of critical states only, but the scaling
behavior of the associated observables can be used to identify the localization characteristics of
all eigenstates in the thermodynamic limit.},
$P_q$ is expected to feature power law dependencies on $l$ and $L$.
The multifractal exponent $\tau_q$ of state $n$ can be defined as
\begin{equation}
\tau_q(n) = \lim_{L/l \to \infty} \tau_q(n,l,L) =  \lim_{L/l \to \infty} \frac {\ln P_q (n,l,L)}{\ln(l/L)}~.
\label{eq:tau_q}
\end{equation}
In our numerical analysis, we consider averages over the disorder distribution and/or
over several states for a given disorder realization. They are obtained by averaging the
moments $P_q (n,l,L) $
\begin{equation}
\tau_q = \lim_{L/l \to \infty} \tau_q(l,L) = \lim_{L/l \to \infty} \frac {\ln \langle P_q (n,l,L)\rangle}{\ln(l/L)}~
\label{eq:tau_q_av}
\end{equation}
where $\langle \ldots \rangle$ denotes the appropriate average.

\subsection{Localization properties: recursive Green function approach}
\label{subsec:recursive}

In addition to the multifractal analysis, we calculate the smallest positive Lyapunov exponent $\gamma$,
characterizing the exponential decay of an eigenstate in a quasi-one-dimensional geometry
by means of the recursive Green function approach \cite{MacKinnon80,MacKinnonKramer83,MacKinnon85}.

We consider a system defined on a quasi-one-dimensional strip of size $N \times L$ sites with $N\gg L$. (Geometrically,
it can be considered a being stack of $N$ layers, each containing $L$ sites.) If the Hamiltonian contains nearest-neighbor
interactions only, the coupling matrices $\mathbf{X}_G$ and $\mathbf{X}_H$ defined in eqs.\ (\ref{eq:X_G}) and (\ref{eq:X_H})
take a block-tridiagonal form
	\begin{eqnarray}
	\mathbf{X}=\begin{pmatrix}
	\mathbf{X}_{1} & \mathbf{T}_{1} &  &  &  \\
	\mathbf{T}_{1}^{T} & \mathbf{X}_{2} & \mathbf{T}_{2} & & \\
	& \mathbf{T}_{2}^{T}  & \mathbf{X}_{3} & \ddots & \\
	&  & \ddots & \ddots & \mathbf{T}_{N-1}\\
	&  &  & \mathbf{T}_{N-1}^{T} &
	\mathbf{X}_{N}
	\end{pmatrix}\;. \label{eq:tridiag_form}
	\end{eqnarray}
The diagonal block $\mathbf{X}_i$ contains the onsite terms as well couplings within the $i$th layer, while the upper diagonal block $\mathbf{T}_i$ contains the interlayer couplings between the layers $i$ and $i+1$.

We consider the Green function $\mathbf{g}(\omega^2)=\lim_{\eta\rightarrow 0} \left[(\omega^2+i\eta)\mathbf{I}-\mathbf{X}\right]^{-1}$
associated with $\mathbf{X}$ at energy $\omega$. Here, $\mathbf{I}$ is the identity matrix and $\eta$ shifts the energy into the complex plane to avoid singularities. We use the tridiagonal form (\ref{eq:tridiag_form}) to calculate the Green function and related physical quantities in a recursive manner. We start with a single layer, and each further iteration step adds a new layer to the stack.
Denoting the Green functions between layers $j$ and $k$ after $N$ iterations as $\mathbf{g}_{j,k}^N$, the recursion can be written as
\begin{eqnarray}
	\mathbf{g}^{N+1}_{1,N+1}&=&\mathbf{g}^{N}_{1,N} \cdot \mathbf{T}_{N} \cdot \mathbf{g}^{N+1}_{N+1,N+1}\;, \\
	\mathbf{g}^{N+1}_{N+1,N+1} &=& \left[(\omega^2+i\eta)\mathbf{I}-\mathbf{X}_{N+1}-
                                    \mathbf{T}_{N}^{T}\cdot\mathbf{g}^{N}_{N,N}\cdot\mathbf{T}_{N} \right]^{-1} ~.
\end{eqnarray}
The iteration is initialized by $\mathbf{g}^1_{11}=\left[(\omega^2+i\eta)\mathbf{I}-\mathbf{X}_1\right]^{-1}$.
In this formalism, the smallest positive Lyapunov exponent (inverse localization length) reads
\begin{eqnarray}
	\gamma(\omega^2,L)=\lim\limits_{N\rightarrow \infty}\frac{1}{2N} \ln|\mathbf{g}^{N}_{1N}|^2\;.\label{eq:lyapunov}
\end{eqnarray}
Localization properties are conveniently extracted from the finite-size scaling behavior
of the dimensionless Lyapunov exponent $\Gamma=\gamma L$.
%Similarly, we also calculate the density of states $\rho(\omega)$. The algorithm, see Ref.\ \cite{MacKinnon85}, calculates
%the density $\tilde \rho(\omega^2)$, which can be transformed to $\rho(\omega)$ via $\rho(\omega)=2\omega\tilde\rho(\omega^2)$.

The matrix elements in $\mathbf{g}^{N}_{1,N}$ usually decay quickly with increasing $N$. To sustain numerical stability
during the iteration, we extract the
leading order of magnitude of the elements after each $\kappa=100$ iterations. Mathematically, this means that
we rewrite the logarithm in eq.\ (\ref{eq:lyapunov}) as
\begin{equation}
	\ln|\mathbf{g}^{N}_{1,N}|^2 = \ln |\mathbf{g}^{\kappa}_{1,\kappa}|^2 +\sum_{b=2}^{N/\kappa} \ln \frac{|\mathbf{g}^{b\kappa}_{1,b\kappa}|^2}{ |\mathbf{g}^{(b-1)\kappa}_{1,(b-1)\kappa}|^2}\;.
\end{equation}

%%%%%%%%%%%%%%%%%%%%%%%%%%%%%%%%%%%%%%%%%%%%%%%%%%%%%%%%%%%%%%%%%%%%%%%%%%%%%%%%%%%%%
\section{Collective excitations in the clean Bose-Hubbard model}
\label{sec:clean}
%%%%%%%%%%%%%%%%%%%%%%%%%%%%%%%%%%%%%%%%%%%%%%%%%%%%%%%%%%%%%%%%%%%%%%%%%%%%%%%%%%%%%

In this section, we apply our approach to the clean Bose-Hubbard model with nearest-neighbor interactions,
\begin{equation}
   H= - J \sum_{\langle ij \rangle} (a^\dagger_i a_j + \mathrm{h.c.}) + \frac{U}{2} \sum_i (n_i-\bar{n})^2~,
   \label{eq:BHM_clean}
\end{equation}
defined on a Bravais lattice of $N$ sites with coordination number (number of nearest neighbors) $z$.
In the first sum, $\langle ij \rangle$ denotes pairs of nearest-neighbor sites. The results reproduce
earlier work in the literature \cite{AltmanAuerbach02,Pekkeretal12}; they are summarized here
to simplify the comparison of the clean and disordered cases.
The Hamiltonian (\ref{eq:BHM_clean}) is a special case of the general Bose-Hubbard model
(\ref{eq:BHM}) with $U_i \equiv U$ and $J_{ij} \equiv J$ if sites $i$ and $j$
are nearest neighbors (but zero otherwise).
After truncation of the local Hilbert spaces, the corresponding pseudo-spin-1 Hamiltonian
reads
\begin{equation}
	H_S = -\frac {\tilde J} 2 \sum_{\langle ij \rangle}  (S^+_i S^-_j + \mathrm{h.c.}) + \frac U 2 \sum_i  (S^z_i)^2~,
\label{eq:H_spin_clean}
\end{equation}
where $\tilde J=\bar n J$.

In a Bravais lattice, all sites are equivalent. This implies that the variational parameters
in the product ansatz (\ref{eq:gs_wavefunction}) for the ground state wave function do not depend
on the lattice site, $\theta_i \equiv \theta$, $\eta_i \equiv \eta$. The variational ground state
energy simplifies to
\begin{eqnarray}
	E_{0}=\langle \Phi_0 | H_s | \Phi_0 \rangle = \frac 1 2 N U \Shsq{} - \frac 1 2 N z \tilde J \Ssq{}
 \label{eq:thermodnamic_groundstate_clean}
\end{eqnarray}
where $N$ is the number of sites.
It is minimized by a mixing angle $\theta$ that fulfills the mean-field equation
$4 z \tilde J \C{} \S{}  = U \S{}$.
A superfluid solution $\C{} = U/U_{c0}$ appears for interactions $U$ below the critical value
$U_{c0}=4z \tilde J$. Its local order parameter $\psi_j = \langle S_j^+\rangle=\Rc{}\S{}$
is uniform in space; it has magnitude $\S{}=\sqrt{1-(U/U_{c0})^2}$ and arbitrary phase $\eta$
(which we set to zero, as before).

We now turn to excitations on top of the mean-field ground state. In the clean case,
the fluctuation Hamiltonians (\ref{eq:H_G}) and (\ref{eq:H_H}) take the form
\begin{eqnarray}
\mathcal{H}_{G} &=& -  \tilde J \Chsq{} \sum_{\langle ij \rangle}  (b_{G,i}+b_{G,i}^\dagger) (b_{G,j}+b_{G,j}^\dagger)
         + \varpi_{G} \sum_{i} n_{G,i} ~, \label{eq:H_G_clean} \\
\mathcal{H}_{H} &=& -  \tilde J \Csq{} \sum_{\langle ij \rangle}  (b_{H,i}+b_{H,i}^\dagger) (b_{H,j}+b_{H,j}^\dagger)
         + \varpi_{H} \sum_{i} n_{H,i} \label{eq:H_H_clean}
\end{eqnarray}
with $\varpi_{G} = (U/2) \Chsq{} + z\tilde J \Ssq{} $ and $\varpi_{H} = (U/2) \C{} + 2 z\tilde J \Ssq{} $.
These Hamiltonians are translationally invariant. After a spatial Fourier transformation,
they can be written as
\begin{equation}
\mathcal{H}_\alpha = \frac{1}{2}\sum_\mathbf{q} \begin{pmatrix}
 b_{\alpha,\mathbf{q}}^\dagger ,
 {b}_{\alpha,-\mathbf{q}}
 \end{pmatrix}
 \begin{pmatrix}
 C_{\alpha,\mathbf{q}} & D_{\alpha,\mathbf{q}} \\
 D_{\alpha,\mathbf{q}} & C_{\alpha,\mathbf{q}}
 \end{pmatrix}
 \begin{pmatrix}
 {b}_{\alpha,\mathbf{q}} \\
 {b}_{\alpha,-\mathbf{q}}^\dagger
 \end{pmatrix}
 +\textrm{const}
\label{eq:H_alpha_clean}
\end{equation}
where $\mathbf{q}$ is the wave vector. The coefficients read
\begin{eqnarray}
C_{G,\mathbf{q}}  &=& \varpi_{G}-D_{G,\mathbf{q}} ~, \qquad D_{G,\mathbf{q}} =  \epsilon(\mathbf{q}) \Chsq{}~, \\
C_{H,\mathbf{q}}  &=& \varpi_{H}-D_{H,\mathbf{q}} ~, \qquad D_{H,\mathbf{q}} =  \epsilon(\mathbf{q}) \Csq{}~.
\end{eqnarray}
Here $\epsilon(\mathbf{q}) = \sum_j \tilde J_{ij} e^{i \mathbf{q} \cdot \mathbf{r}_{ij} } $ is the dispersion of the lattice.
For a square lattice with nearest-neighbor interactions, it is given by
$\epsilon(\mathbf{q}) = 2 \tilde J \cos(q_x) + 2 \tilde J \cos(q_y)$.

\begin{figure}[t]
\includegraphics[width=\textwidth]{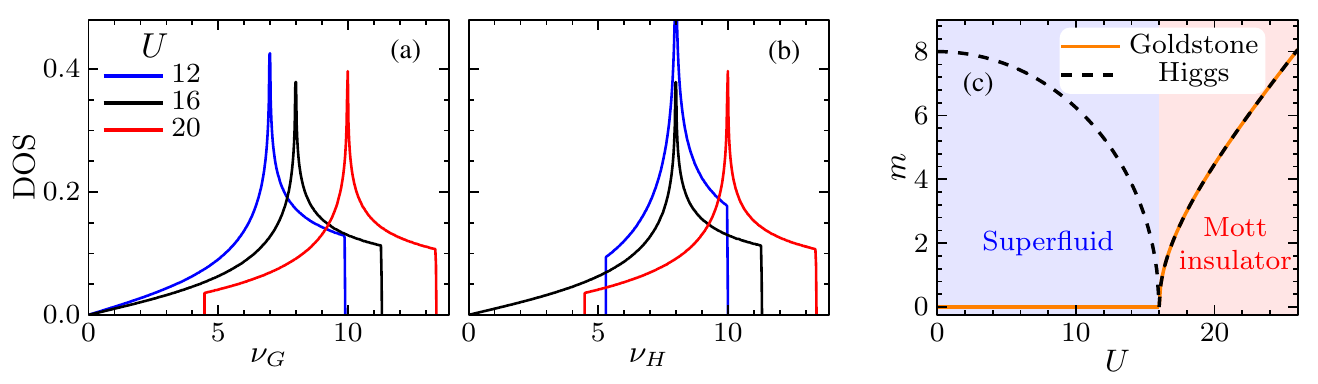}
\caption{Eigenfrequency spectra (densities of states) of the collective Goldstone (a) and Higgs (b)
modes for the clean square lattice Bose-Hubbard model for three values of $U$: superfluid ($U=12$),
critical point ($U=16$), Mott phase ($U=20$). ($\tilde J$ is set to unity.)
(c)
 The masses (energy gaps) $m_G$  and $m_H$ of the
Goldstone and Higgs collective modes. }
\label{fig:DOS_clean}
\end{figure}

The Hamiltonians (\ref{eq:H_alpha_clean}) can be solved by applying a (unimodal) Bogoliubov transformation to each
$(\mathbf{q}, -\mathbf{q})$ pair. This yields
\begin{equation}
\mathcal{H}_\alpha = \sum_\mathbf{q} \nu_{\alpha,\mathbf{q}} d_{\alpha,\mathbf{q}}^\dagger d_{\alpha,\mathbf{q}}
\end{equation}
with the excitation eigenenergies given by
\begin{eqnarray}
\nu_{G,\mathbf{q}}^2 &=&  [\varpi_G -  \epsilon(\mathbf{q}) \Chsq{} ]^2  - [ \epsilon(\mathbf{q}) \Chsq{}]^2 ~,   \\
\nu_{H,\mathbf{q}}^2 &=&  [\varpi_H -  \epsilon(\mathbf{q}) \Csq{} ]^2  - [ \epsilon(\mathbf{q}) \Csq{}]^2 ~.
\end{eqnarray}
The same result can also be obtained by using the first-quantization framework of Sec.\ \ref{sec:bogoliubov},
i.e., by transforming the fluctuation Hamiltonians (\ref{eq:H_G_clean}) and (\ref{eq:H_H_clean}) to position and momentum
representation and diagonalizing the resulting coupling matrix via Fourier transformation.
Figures \ref{fig:DOS_clean}(a) and \ref{fig:DOS_clean}(b) shows the resulting densities of states of the collective excitations.

The mass (or energy gap) $m_\alpha$ of the Goldstone and Higgs modes is given by the lowest excitation energy
for a given set of parameters,
i.e.,  $m_\alpha = \nu_{\alpha,\mathbf{q}=0}$. Using $\epsilon(\mathbf{q}=0)=z \tilde J$, we therefore obtain
$m_G^2 = m_H^2 = (U/4)(U-U_{c0})$  in the Mott insulating phase ($U>U_{c0}$).
In the superfluid phase ($U<U_{c0}$), the gap of Goldstone mode vanishes, $m_G^2=0$, whereas the gap
of the Higgs mode reads $m_H^2=(1/4)(U_{c0}^2 -U^2)$. Thus the Higgs mode is gapless at the critical point $U=U_{c0}$
only. The dependence of the masses on $U$ is illustrated in Fig.\ \ref{fig:DOS_clean}(c).

%%%%%%%%%%%%%%%%%%%%%%%%%%%%%%%%%%%%%%%%%%%%%%%%%%%%%%%%%%%%%%%%%%%%%%%%%%%%%%%%%%%%%
\section{Simulations and results}
\label{sec:simulations}
%%%%%%%%%%%%%%%%%%%%%%%%%%%%%%%%%%%%%%%%%%%%%%%%%%%%%%%%%%%%%%%%%%%%%%%%%%%%%%%%%%%%%
\subsection{Overview}
%%%%%%%%%%%%%%%%%%%%%%%%%%%%%%%%%%%%%%%%%%%%%%%%%%%%%%%%%%%%%%%%%%%%%%%%%%%%%%%%%%%%%

We now turn to the properties of the disordered Bose-Hubbard model for which the mean-field
theory developed in Sec.\ \ref{sec:mftheory} needs to be solved by means of computer simulations.
Specifically, we focus on a square-lattice Bose-Hubbard model with nearest neighbor hopping.
The quantum phase transition between superfluid and Mott insulator is tuned by varying the
Hubbard interaction $U$ whereas
the scaled hopping amplitude is set to unity,  $\tilde J=1$, fixing the energy scale.
As described in Sec.\ \ref{sec:BHM}, we consider two types of (particle-hole-symmetry
preserving) disorder, site dilution and random Hubbard interactions $U_i$.

In the case of site dilution, we employ dilution values (vacancy probabilities)
$p=0$, 1/8, 1/5, 1/4, and 1/3. For comparison, the site percolation threshold for the
square lattice is $p_c=0.407253$ \cite{StaufferAharony_book91}. For dilutions
$p>p_c$, the lattice consists of disconnected finite-size clusters only that do not support
long-range superfluid order. For dilutions $p<p_c$, an ``infinite'' cluster that spans
the entire sample coexists with disconnected finite-size clusters. Since we are interested
in long-range ordered states, we only consider the infinite cluster and neglect the
finite clusters.
In the case of random-$U$ disorder, the lattice is undiluted but the local Hubbard interactions
$U_i$ are drawn from the distribution (\ref{eq:U_distrib}). We consider disorder strengths
$r= 0.5$, 1.0 and 1.5.

Most calculations are performed on lattices of $L \times L$ sites with linear sizes
up to $L=256$. For a few calculations of excitations at higher energies, we
employ linear sizes as large as $L=1536$.
In addition, we analyze strips of up to $128 \times 10^6$ sites within the recursive
Green function approach of Sec.\ \ref{subsec:recursive}. All results are averaged over a large
number of disorder realizations.

The first step in our approach is the numerical solution of the mean-field equations
(\ref{eq:sf_groundstate_condition}) which constitute a large system of coupled nonlinear
equations. We implement two different numerical algorithms to solve these equations
efficiently and accurately, a simple iterative method and a gradient descent method.
In \ref{app:mf_numerics}, we describe these methods in more detail and discuss
their accuracy and numerical stability.

%%%%%%%%%%%%%%%%%%%%%%%%%%%%%%%%%%%%%%%%%%%%%%%%%%%%%%%%%%%%%%%%%%%%%%%%%%%%%%%%%%%%%
\subsection{Mean-field ground state}
\label{subsec:mf_GS_results}
%%%%%%%%%%%%%%%%%%%%%%%%%%%%%%%%%%%%%%%%%%%%%%%%%%%%%%%%%%%%%%%%%%%%%%%%%%%%%%%%%%%%%

\begin{figure}[t]
\includegraphics[width=6.9cm]{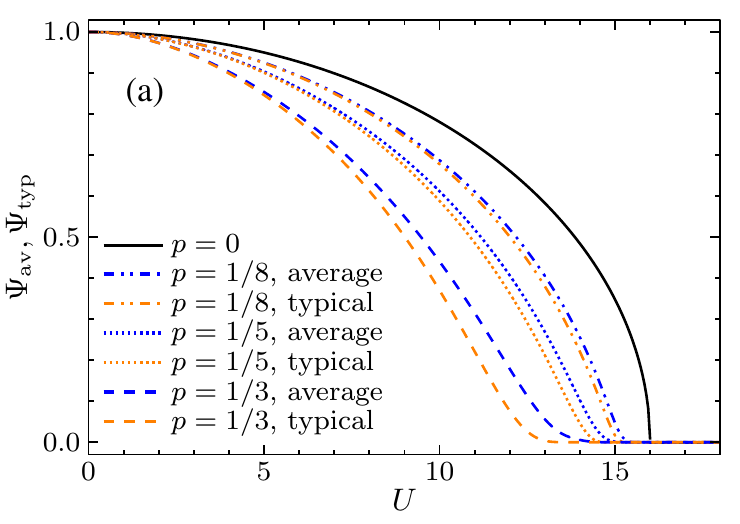}~ \includegraphics[width=6.2cm]{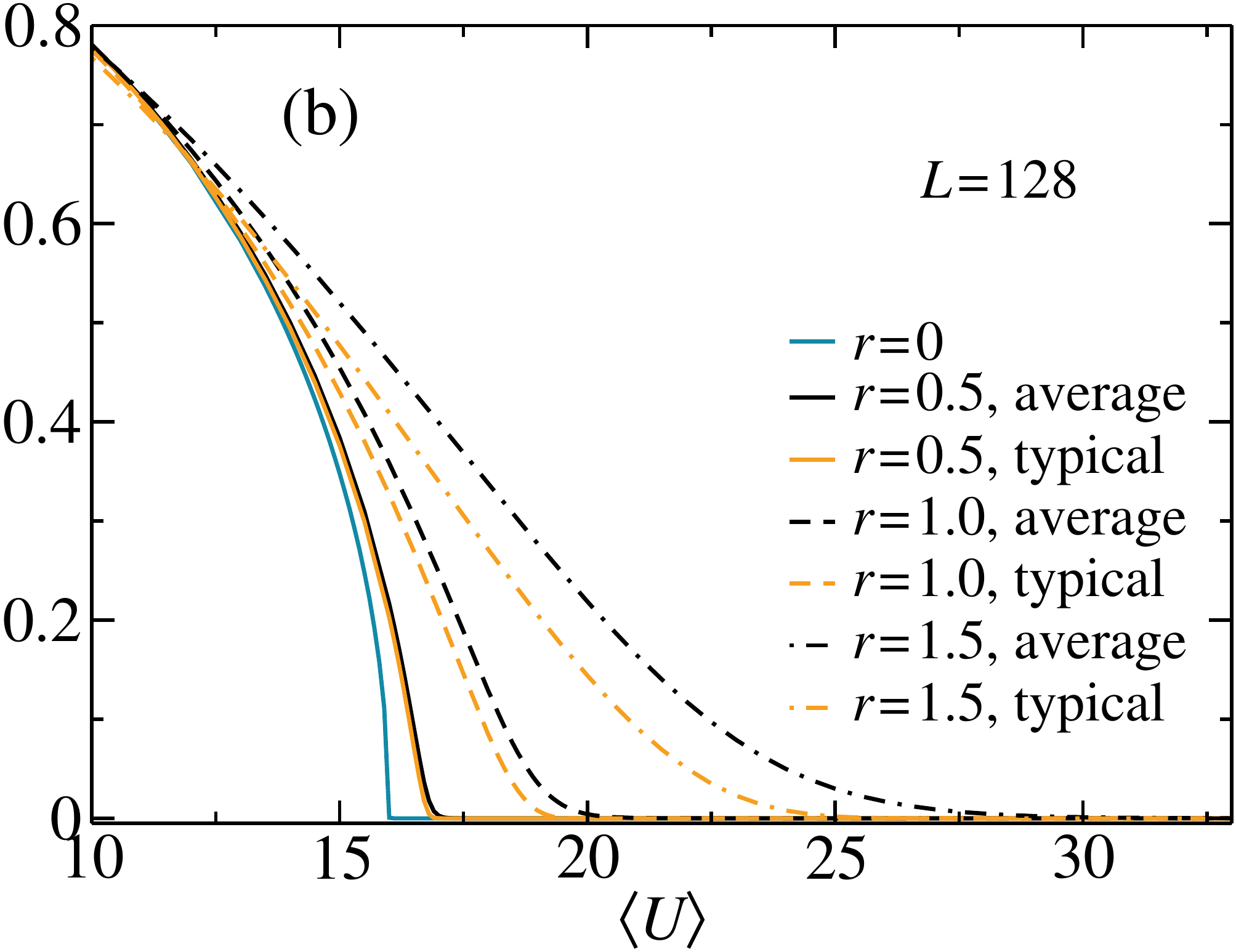}
\caption{Average and typical order parameters $\Psi_\textrm{av}$ and $\Psi_\textrm{typ}$ vs.\
Hubbard interaction $U$. (a) Site-diluted systems of dilutions $p=0, 1/8, 1/5$, and 1/3.
(b) Random-$U$ disorder with disorder strength $r=0.5, 1$, and 1.5. The data are based
on (arithmetic) averages of $\Psi_\textrm{av}$ and $\Psi_\textrm{typ}$ over 1000 disorder
realizations for square lattices of linear size $L=128$. The statistical errors are below
the line thickness.}
\label{fig:OP_av_typ}
\end{figure}

We now discuss the results of our computer simulations, starting with the properties of the
mean-field ground state. To characterize the superfluid order, we compute the average and typical
order parameters for a given sample (disorder realization), defined as the arithmetic and
geometric means, respectively, of the local order parameters $\psi_j = \sin(\theta_j)$,
\begin{equation}
\Psi_\textrm{av} = \frac 1 N \sum_j \psi_j \quad,  \qquad \Psi_\textrm{typ} = \exp \left ( \frac 1 N \sum_j \ln(\psi_j) \right)~.
\label{eq:OP_defs}
\end{equation}
Figure \ref{fig:OP_av_typ} presents the dependence of these quantities, averaged over a large number of
disorder realizations,\footnote{We have employed an arithmetic average over the disorder configurations;
using a geometric average gives essentially the same results. This indicates that $\Psi_\textrm{av}$ and
$\Psi_\textrm{typ}$ are already self-averaging for the system sizes considered.}
on the Hubbard interaction for several strengths of both disorder types.
In the clean case (included in Fig.\ \ref{fig:OP_av_typ}(a) as dilution $p=0$
and in Fig.\ \ref{fig:OP_av_typ}(b) as $r=0$),
the average and typical order parameter coincide
because all local order parameters $\psi_j$ are identical. They follow the typical
mean-field behavior  $\Psi_\textrm{av} = \Psi_\textrm{typ} = \sqrt{1-(U/U_{c0})^2}$
for $U<U_{c0}=16$  and $\Psi_\textrm{av} = \Psi_\textrm{typ} =0$ for $U>U_{c0}$,
as derived in Sec.\ \ref{sec:clean}.

Figure \ref{fig:OP_av_typ}(a) demonstrates that the superfluid order is suppressed
with increasing dilution, as expected, because the missing neighbors lead to an overall
reduction of the coupling strength. Close to the onset of superfluidity, the typical order
parameter is significantly smaller than the average one, indicating that rare superfluid
islands (puddles) coexist with insulating regions. These rare puddles are also responsible
for the pronounced tails of the order parameter curves towards large $U$. In an
infinite system, impurity-free regions of arbitrary size exist with exponentially small
(in their size) but nonzero probability. As the largest of these regions develop superfluid
order for $U$ values right below the clean critical point $U_{c0}=16$, the exponential tails
of the order parameter curves stretch all the way to the clean critical point in the
thermodynamic limit.
It must be emphasized that these tails are artifacts of the mean-field theory which is unable
to describe fluctuations of the superfluid order. In reality, the superfluid order on
isolated rare regions is not static and thus does not contribute to the order parameter.
The rare regions instead fluctuate, as expected in a quantum Griffiths phase \cite{Vojta06,Vojta10}.
Indeed, quantum Monte Carlo simulations do not show exponential tails in the order parameter
curves but sharp power-law singularities associated with a conventional critical point
at $U_c(p) < U_{c0}$  \cite{Vojtaetal16,LerchVojta19}. Effectively, the mean-field approximation
(incorrectly) replaces the quantum Griffiths phase by the tail of a smeared quantum phase
transition \cite{Vojta03a,Vojta03b}. This also implies that the exact location of the quantum phase
transition cannot be determined within the mean-field theory.
Deeper inside the ordered phase where superfluidity is not restricted to rare
puddles but becomes more homogeneous, the mean-field description becomes qualitatively correct.

Figure \ref{fig:OP_av_typ}(b) shows the order parameter curves for the case of random-$U$
disorder. Interestingly, the onset of superfluidity shifts to larger (average) Hubbard
interaction $\langle U \rangle$ with increasing disorder strength. This is caused by spatial
regions in which the majority of the local Hubbard interactions $U_i$ are below the average
$\langle U \rangle$. These regions are locally in the superfluid phase before $\langle U \rangle$
reaches the clean critical value $U_{c0}$. Analogous to the case of dilution disorder,
the order parameter curves develop spurious exponential tails towards large $U$. In the
thermodynamic limit, these tails terminate at $U_{c0}/(1-r/2)$.

\begin{figure}[t]
\includegraphics[width=\textwidth]{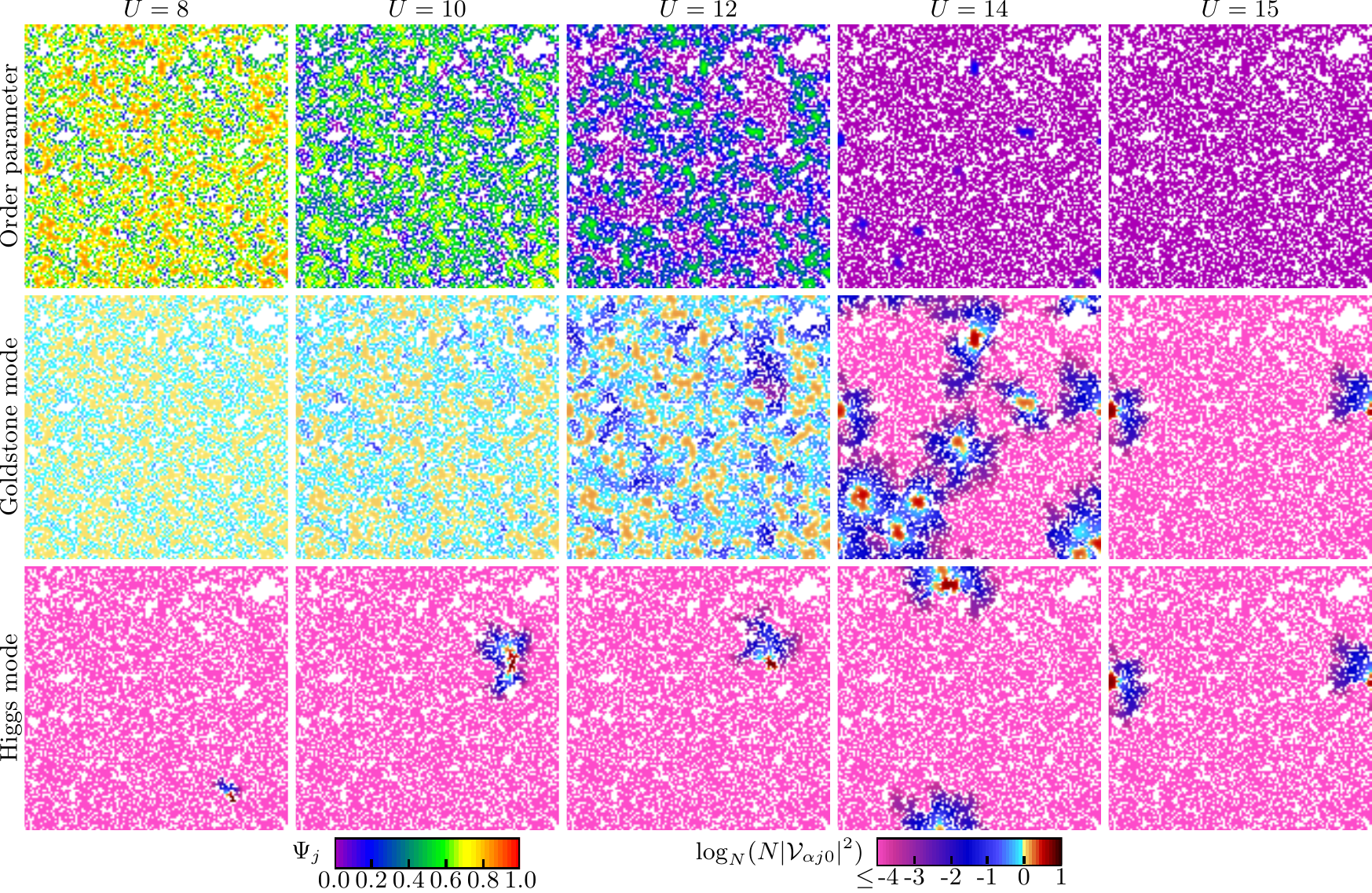}
\caption{Local order parameters $\psi_j$, and eigenstate wavefunctions
$\mathcal{V}_{Gj0}$ and $\mathcal{V}_{Hj0}$ of the lowest Goldstone and Higgs excitations of a single
disorder realization (diluted lattice with $L=128$, $p=1/3$, and different $U$).
White sites represent vacancies or sites of disconnected finite-size clusters that have been neglected.}
\label{fig:LocalOPandGS}
\end{figure}

To further elucidate the inhomogeneous mean-field ground state, the first row of Fig.\ \ref{fig:LocalOPandGS}
presents heat maps of the local order parameter $\psi_j$ for a single disorder realization
of a system with dilution $p=1/3$.
The figure illustrates that the superfluid order consists of puddles embedded in an insulating bulk
for $U$ values close to the quantum phase transition (i.e., in the tail of the superfluid phase).
In contrast, the order parameter is only moderately inhomogeneous at lower $U$, deeper in the
superfluid phase. An analogous plot of the local order parameter for the case of random-$U$ disorder
is shown in Fig.\ \ref{fig:LocalOPandGS_randomU} in \ref{app:randomU}.

%%%%%%%%%%%%%%%%%%%%%%%%%%%%%%%%%%%%%%%%%%%%%%%%%%%%%%%%%%%%%%%%%%%%%%%%%%%%%%%%%%%%%
\subsection{Excitation spectrum}
\label{subsec:spectrum_results}
%%%%%%%%%%%%%%%%%%%%%%%%%%%%%%%%%%%%%%%%%%%%%%%%%%%%%%%%%%%%%%%%%%%%%%%%%%%%%%%%%%%%%

\begin{figure}[t]
\includegraphics[width=\textwidth]{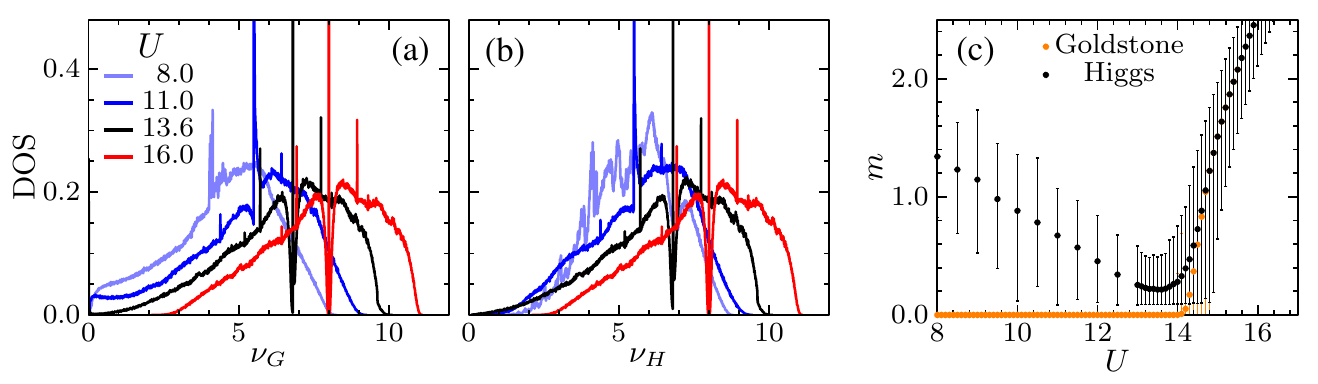}
\caption{Eigenfrequency spectra (density of states) of the Goldstone (a) and Higgs (b)
modes for dilution $p=1/3$ and four interaction values,
$U=8$ (superfluid phase), 11 and 13.6 (close to the quantum phase transition), and
16 (Mott-insulating phase).
The data are averages over 384 disorder realizations of square lattices of linear size $L=128$.
(c) Masses (energy gaps) $m_G$  and $m_H$ of the
Goldstone and Higgs modes. The dots show the average over 1000 realizations,
the bars show the spread of observed values.}
\label{fig:DOS_diluted}
\end{figure}

We now turn to the excitations on top of the mean-field ground state, starting with a discussion of
the excitation spectrum, i.e., the eigenvalues of the fluctuation Hamiltonians $\mathcal{H}_{G}$ and
$\mathcal{H}_{H}$ given in eqs.\ (\ref{eq:H_G}) and (\ref{eq:H_H}), respectively.
Figures \ref{fig:DOS_diluted}(a)  and \ref{fig:DOS_diluted}(b)
present the densities of state for the Goldstone and Higgs excitations
for the case of dilution disorder with $p=1/3$ for several values of $U$.
The qualitative behavior
is analogous to the clean case shown in Fig.\ \ref{fig:DOS_clean}. In the superfluid phase,
the Higgs mode has a nonzero energy gap (mass) $m_H$ whereas the energy gap $m_G$ of the Goldstone mode
vanishes identically in agreement with Goldstone's theorem.  In the Mott-insulating phase, the two excitation sectors
are degenerate and gapped. The $U$-dependence of both energy gaps is further illustrated in Fig.\
\ref{fig:DOS_diluted}(c). As expected, the Higgs mode softens close to the superfluid-Mott insulator
transition, i.e., its mass approaches zero. In contrast
to the clean case shown in Fig.\ \ref{fig:DOS_clean}, the (average) Higgs mass does not appear to reach exactly
the value zero at the transition point. This can be attributed to finite-size effects, as each
finite-size sample (disorder realization)
has a slightly different critical $U$ for which a superfluid solution first appears. Averaging the masses
thus smears the gap. This effect is exacerbated by the artifacts of the mean-field solution
discussed in Sec.\ \ref{subsec:mf_GS_results}. The mean-field solution allows static superfluid
order to appear on isolated rare regions, greatly increasing the variations of the critical $U$
between (finite-size) disorder configurations, and shifting their values towards the clean $U_{c0}=16$.

At higher energies, the densities of state of both modes show sharp features. They are caused by the discrete
nature of the dilution disorder, i.e., they stem from small finite-size clusters of sites which
support excitations of fixed energies.

\begin{figure}[b]
\includegraphics[width=4.6cm]{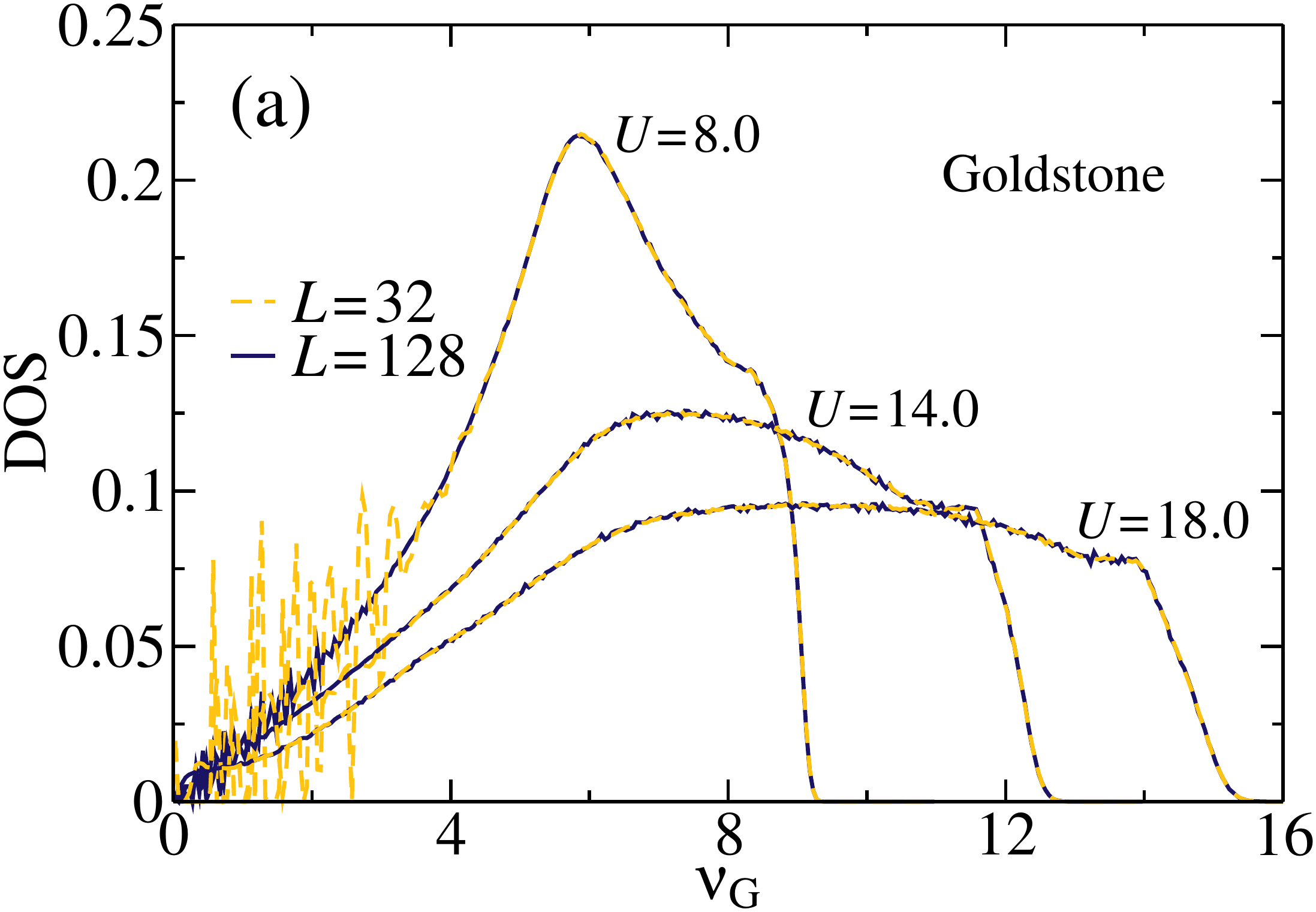}~\includegraphics[width=4.5cm]{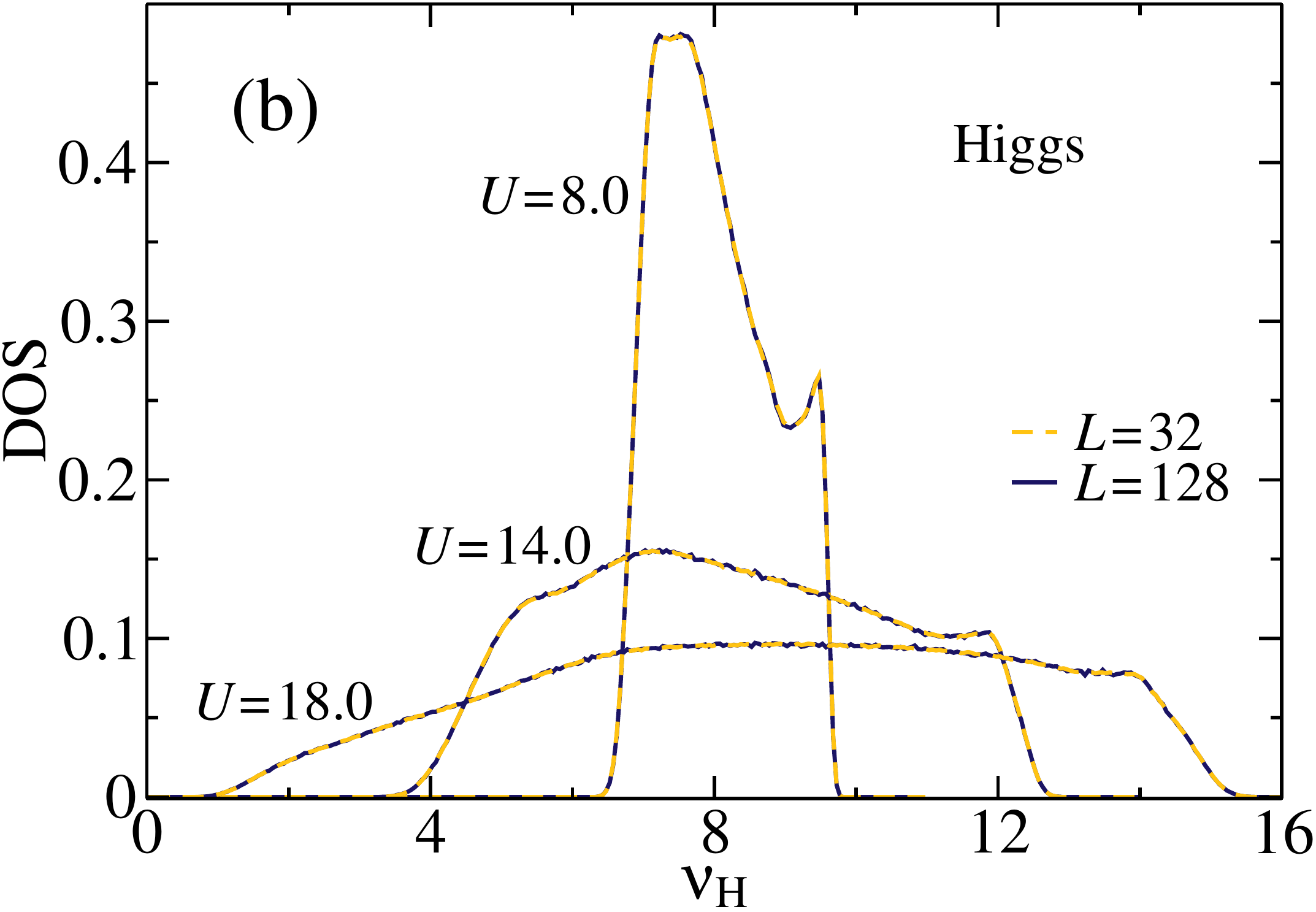}~
  \includegraphics[width=4.2cm]{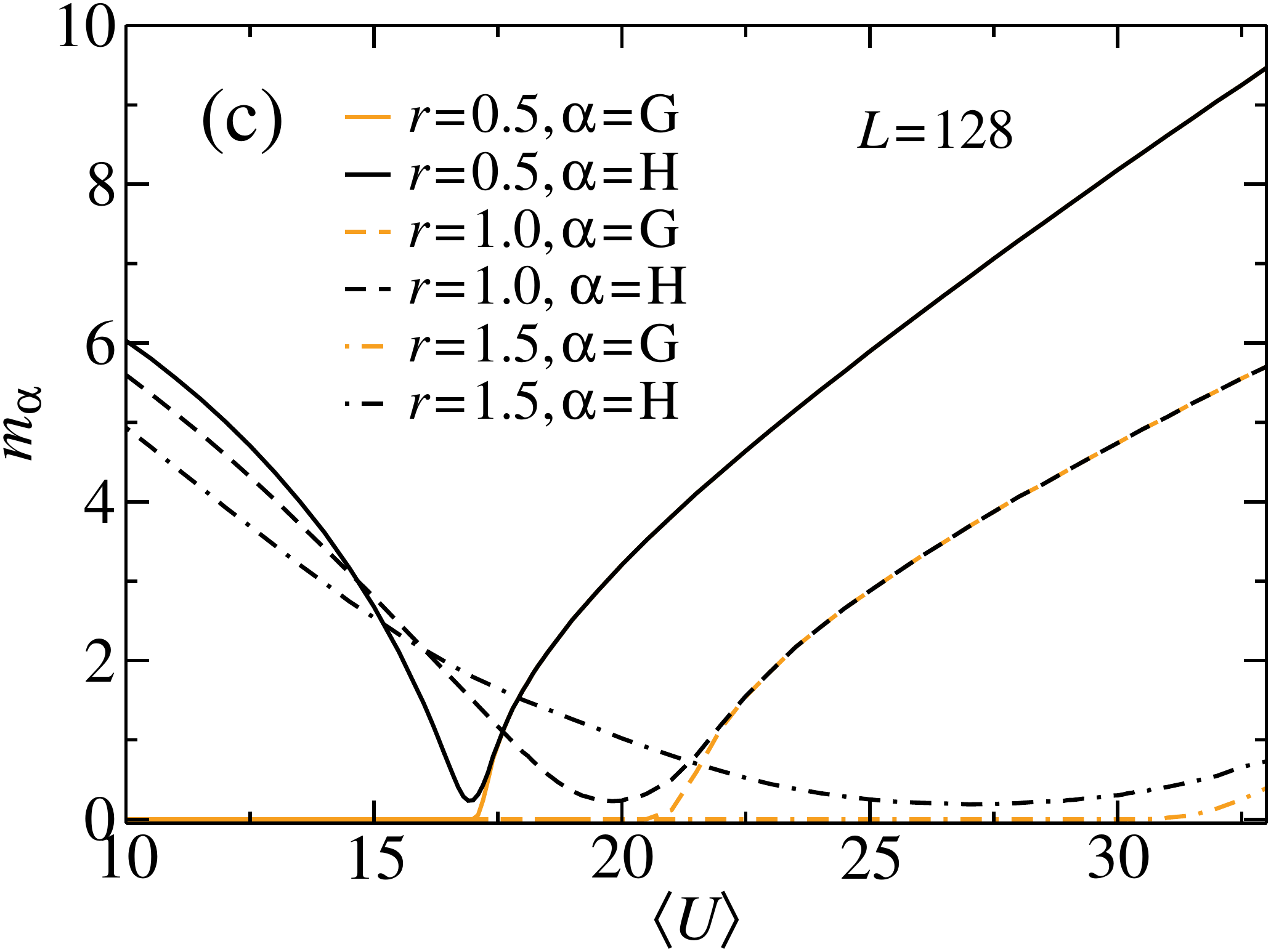}
\caption{Eigenfrequency spectra (density of states) of the Goldstone (a) and Higgs (b)
modes for three values of $U$: $U=8$, and $U=14$ in the superfluid phase as well as $U=18$ close
to the quantum phase transition. (Square lattices of linear sizes $L=32$ and 128,
 random-$U$ disorder with $r=1.0$.)
(c) Masses (energy gaps) $m_G$  and $m_H$ of the
Goldstone and Higgs modes for several disorder strengths $r$.
All data are averages over 1000 disorder realizations. }
\label{fig:DOS_randomU}
\end{figure}

Systems with random-$U$ disorder feature analogous behavior, as can be seen in Fig.\
\ref{fig:DOS_randomU}.
Note that their densities of state do not show the sharp features at higher energies present
for dilution disorder because the random-$U$ distribution is continuous.

%%%%%%%%%%%%%%%%%%%%%%%%%%%%%%%%%%%%%%%%%%%%%%%%%%%%%%%%%%%%%%%%%%%%%%%%%%%%%%%%%%%%%
\subsection{Localization properties of lowest Goldstone and Higgs excitations }
\label{subsec:lowest_excitations_results}
%%%%%%%%%%%%%%%%%%%%%%%%%%%%%%%%%%%%%%%%%%%%%%%%%%%%%%%%%%%%%%%%%%%%%%%%%%%%%%%%%%%%%

After having discussed the energy spectrum of the excitations, we now consider the eigenstates.
The present section focuses on the lowest-energy excitations in both the Goldstone and Higgs channels.
Figure \ref{fig:LocalOPandGS} visualizes examples of their eigenstates for a diluted lattice with $p=1/3$ and several $U$.
Examples of the lowest-energy Goldstone and Higgs eigenstates in a system with random-$U$ disorder are shown in
the appendix in Fig.\ \ref{fig:LocalOPandGS_randomU}. Clearly, the eigenstates feature nontrivial spatial localization
behavior that depends on the channel and varies with $U$.

\begin{figure}[t]
\includegraphics[width=6.6cm]{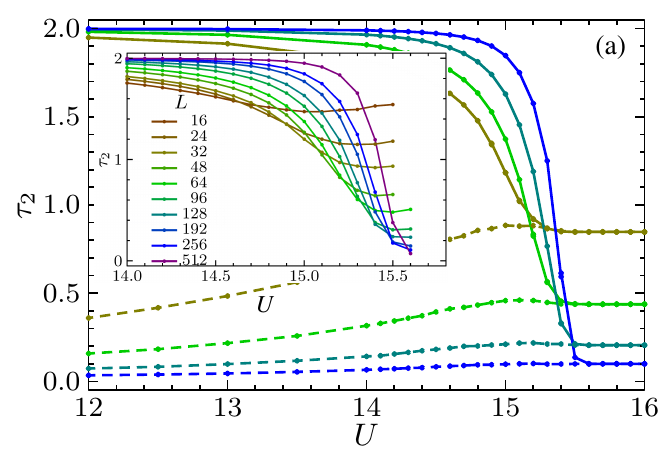} ~ \includegraphics[width=6.6cm]{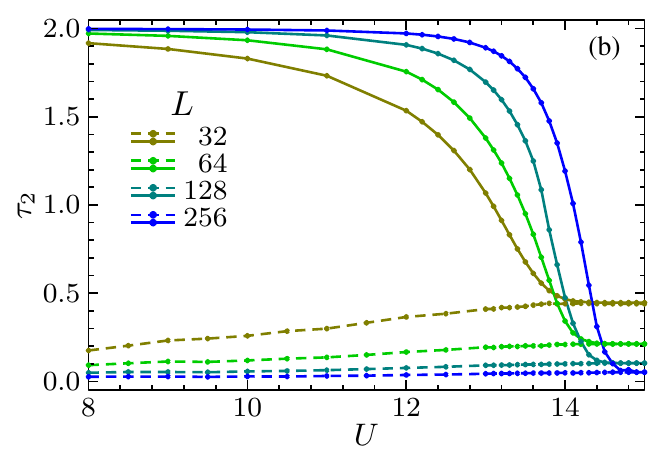} \\
\includegraphics[width=6.5cm]{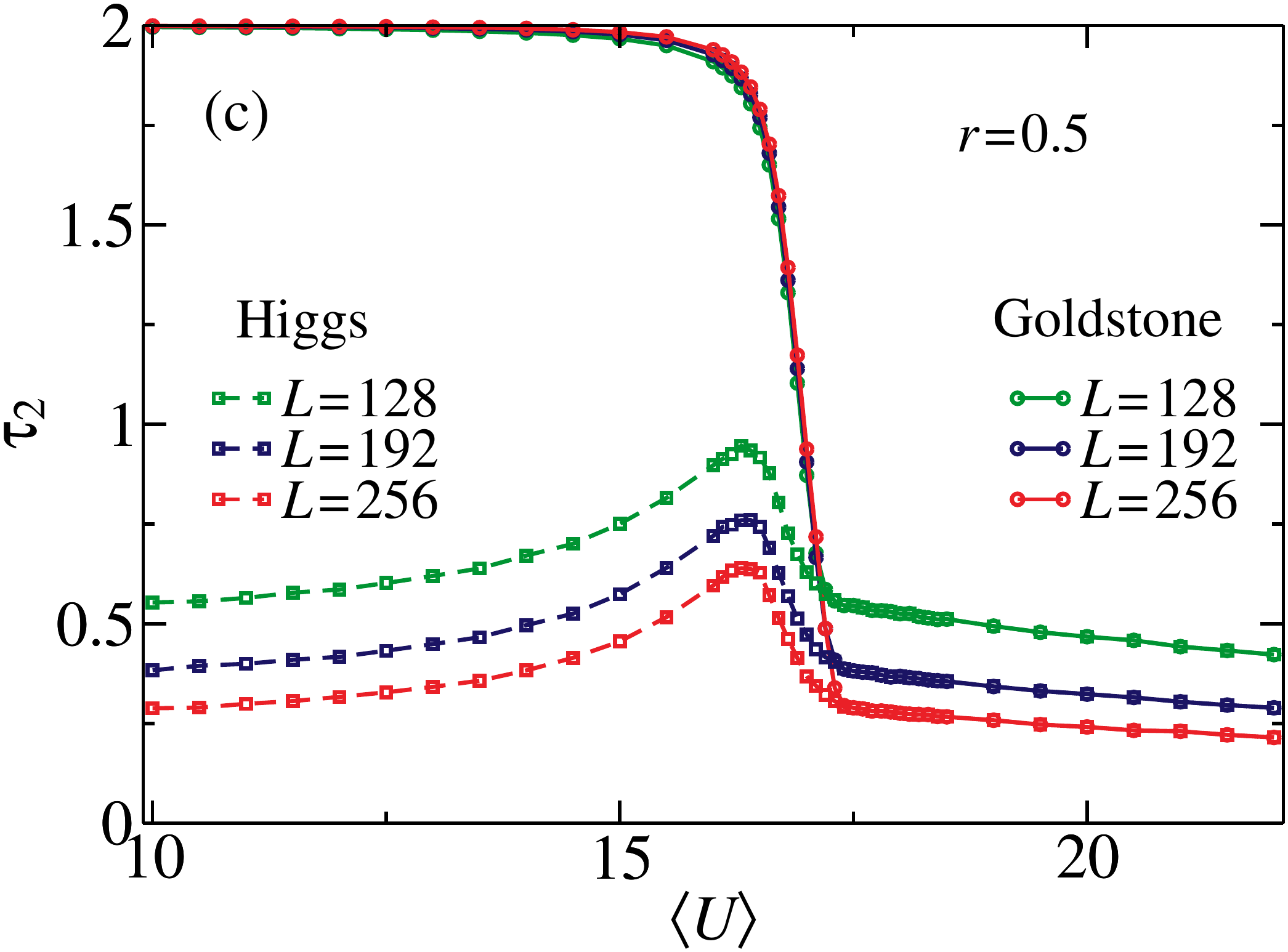}~~~ \includegraphics[width=6.5cm]{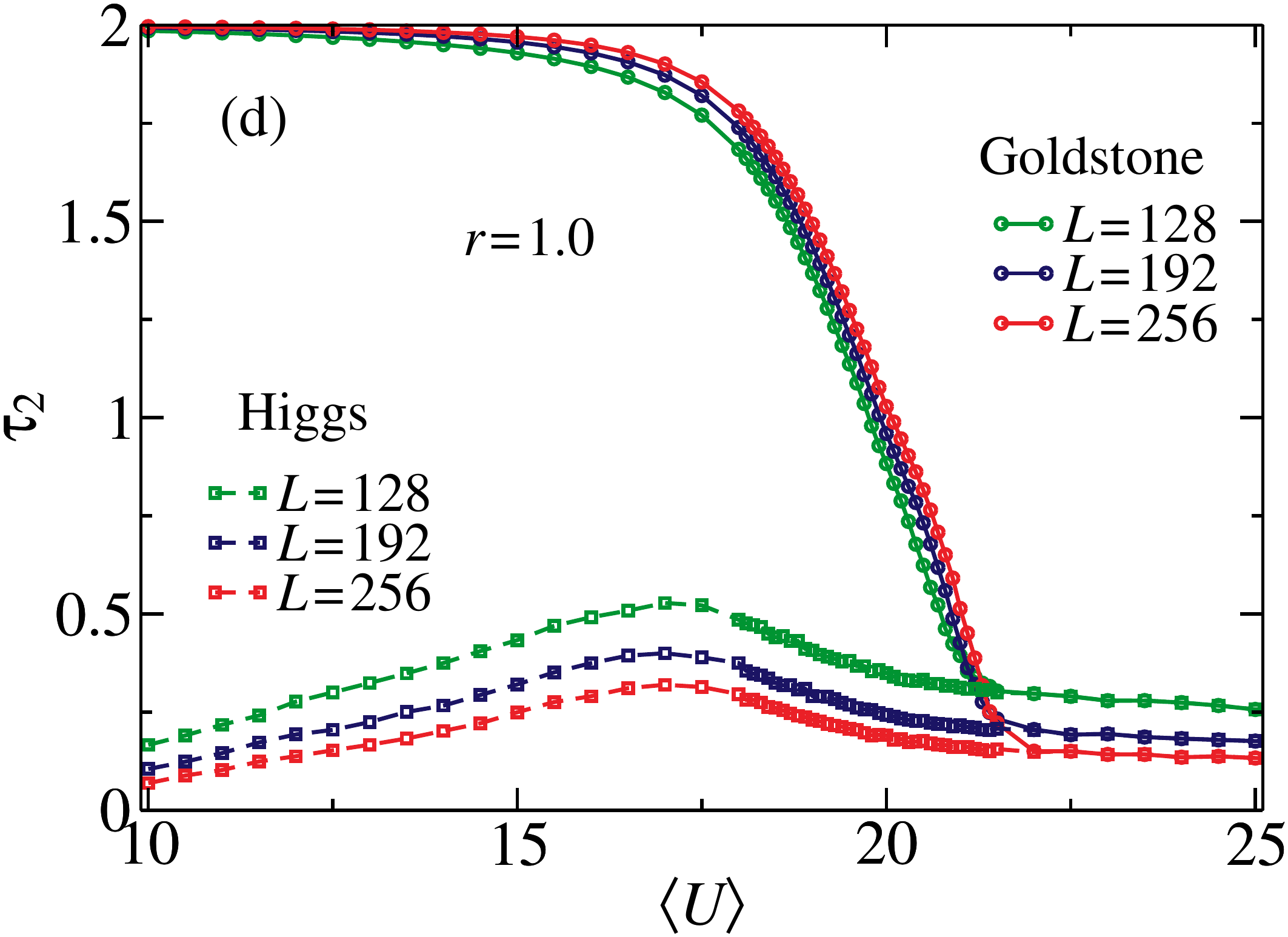}
\caption{Generalized dimension $\tau_2$ of the lowest-energy Goldstone (solid lines) and Higgs
(dashed lines) excitations
vs.\ interaction $U$ for several system sizes $L$. (a) dilution $p=1/8$, (b) dilution $p=1/3$,
(c) random-$U$ disorder of strengths $r=0.5$, and (d) random-$U$ disorder of strengths $r=1.0$.
The box size $l$ is chosen according to $L/l=8$.
The data are averages over 1000 disorder realizations. Statistical errors are smaller than the symbol size.
The inset of panel (a) shows a magnification of the transition region, using the analytic expression
(\ref{eq:HandySolution}) for the lowest Goldstone excitation.}
\label{fig:tau2_lowest}
\end{figure}

To analyze the localization properties quantitatively, we compute the generalized dimension $\tau_2$
of the lowest Goldstone and Higgs excitations, as defined in eq.\ (\ref{eq:tau_q_av}).
The dependence of $\tau_2$ on the Hubbard interaction $U$ is shown in Fig.\ \ref{fig:tau2_lowest}
for both types of disorder and several disorder strengths.
All cases feature the same qualitative behavior. In the Mott-insulating phase (large $U$), both excitations are
degenerate. The $\tau_2$ values rapidly decrease towards zero with increasing system size, indicating strong
localization.

As the system enters the superfluid phase with decreasing $U$, the two excitation branches evolve in
qualitatively different ways.
The lowest Higgs excitation, is localized in the superfluid phase just
as in the insulating phase.  For strong disorder, the degree of localization even seems to increase
in the superfluid phase as $\tau_2$ further decreases.
The lowest Goldstone excitation, in contrast, undergoes a striking delocalization transition
upon entering the superfluid phase. Its $\tau_2$ value rapidly increases with decreasing $U$.
Importantly, the system-size dependence of $\tau_2$ also changes sign, it now increases with increasing $L$
towards the value $\tau_2=2$ characteristic of an extended state.

In fact one can show analytically that the lowest Goldstone excitation is extended over
the entire sample if the system features superfluid long-range order. This can be demonstrated as follows.
According to Goldstone's theorem, the lowest eigenstate of the Goldstone Hamiltonian $\mathcal{H}_G$ must
have zero energy, $\nu_{G,0}= 0$, in the superfluid phase because the superfluid ground state spontaneously
breaks the $U(1)$ order-parameter symmetry. (We have verified this numerically for all samples.)
For this state, the corresponding eigenvalue problem (\ref{eq:EV_problem_first})
simplifies to a system of linear equations,
\begin{equation}
\sum_{k} X_{G,jk} \mathcal{V}_{G, k0} = \nu_{G,0}^2 \mathcal{V}_{G,j0}=0 ~.
\label{S_eq_GS}
\end{equation}
 A non-trivial solution of this system is given by
\begin{eqnarray}
	\mathcal{V}_{G,j0} = \varUpsilon \frac{\sin(\theta_j/2)} {\sqrt{\varpi_{G,j}}}
\label{eq:HandySolution}
\end{eqnarray}
as can be easily checked by inserting it back into the system (\ref{S_eq_GS}). Here,  $\varUpsilon$ is a normalization constant.
Thus, the lowest Goldstone eigenstate depends on the order parameter $\sin (\theta_j)$ and local interactions
(via $\varpi_{G,j}$) only. The denominator of the solution (\ref{eq:HandySolution}) is bounded from both below and above.
(For dilution disorder, $\varpi_{G,j} \ge U /4 $ and $\varpi_{G,j} \le U /2 + 4\tilde J$.)
Consequently, the localization character of $\mathcal{V}_{G,j0}$ is governed by that of the order parameter.
A long-range ordered superfluid state features a nonzero macroscopic order parameter $\Psi_\textrm{av}$ in the thermodynamic limit.
This implies either a more-or-less homogeneous superfluid or at least a nonzero density of superfluid puddles. According to Eq.\ (\ref{eq:HandySolution}), this means that the wave function of the lowest Goldstone excitation is nonzero on a finite fraction of the lattice sites, i.e., it is extended.\footnote{In the Mott-insulating phase, the local order parameters $\psi_j= \sin(\theta_j)$ vanishes on all lattice sites.
Thus, the state (\ref{eq:HandySolution}) is not normalizable, indicating the absence of a zero-energy mode.}

For any given disorder realization, the two excitation channels are degenerate as long as all local
order parameters vanish. This implies that the lowest excitations in both sectors will be extended
right at the critical $U$ for which a superfluid solution first appears. This is reflected in the increase
of the Higgs $\tau_2$ close to the quantum phase transition visible for weaker disorder in Figs.\ \ref{fig:tau2_lowest}
(c) and (d). As each finite-size sample (disorder realization) has a slightly different critical $U$,
the ensemble average masks the diverging Higgs localization length right at the critical $U$.

\begin{figure}[t]
\includegraphics[width=6.5cm]{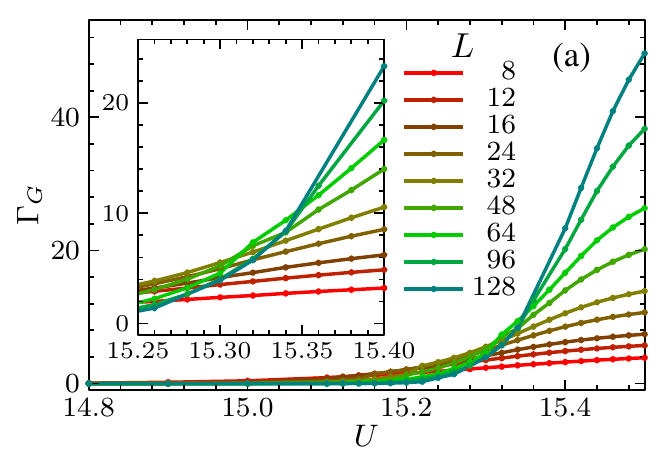}~
\includegraphics[width=6.5cm]{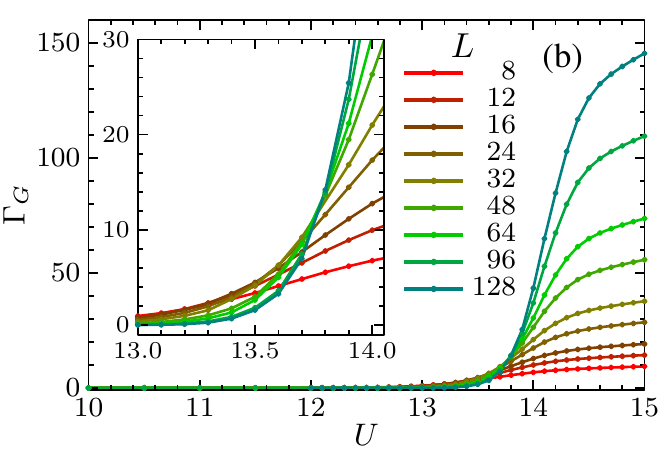}
\caption{Dimensionless Lyapunov exponent of the lowest-energy Goldstone excitation
vs.\ interaction $U$ for several strip width $L$. (a) dilution $p=1/8$. (b) dilution $p=1/3$.
The data are averages over 12 strips of size $L \times 10^6$ sites.
Statistical errors are smaller than the symbol size. The insets show magnifications of the crossing regions. }
\label{fig:Gamma_G}
\end{figure}

The delocalization transition of the lowest Goldstone excitation as a function of $U$ is also observed
in the dimensionless Lyapunov exponent $\Gamma_G$ calculated from quasi-one-dimensional (strip) samples,
as described in Sec.\ \ref{subsec:recursive}.
Figure \ref{fig:Gamma_G} presents the $U$-dependence of the dimensionless Lyapunov exponents for diluted lattices
with $p=1/8$ and $1/3$.
For both dilutions, $\Gamma_G$ rapidly increases with increasing $L$ in the Mott-insulating phase, indicating
strong localization. In the superfluid phase $\Gamma_G$ is small and decreases with increasing $L$, as expected
for spatially extended states.

Our numerical results for the generalized dimension $\tau_2$ and the dimensionless Lyapunov exponent $\Gamma$
as well as the analytical construction (\ref{eq:HandySolution}) demonstrate that the location of the delocalization
transition of the lowest Goldstone excitation as a function of $U$ is tied to that of the (thermodynamic)
quantum phase transition between Mott insulator
and superfluid. A numerical verification that the two transitions coincide
suffers from the same mean-field artifacts already discussed in Secs.\ \ref{subsec:mf_GS_results} and
\ref{subsec:spectrum_results}: As mean-field theory produces a spurious tail of the ordered phase, the seeming
transition point moves to larger $U$ with increasing system size, reaching the clean critical value $U_{c0}$ in
the thermodynamic limit in the case of dilution disorder.\footnote{For random-$U$ disorder, the tail
is expected to stretch to  $U_{c0}/(1-r/2)$.} This effect can actually be observed in our numerical data.
The insets of Figs.\ \ref{fig:tau2_lowest}(a), \ref{fig:Gamma_G}(a), and \ref{fig:Gamma_G}(b) show
that the crossing points of the $\tau_2$ and $\Gamma_G$ curves with consecutive system sizes $L$
move towards larger $U$ and towards the localized limit with increasing $L$.

We emphasize that the qualitative features of the localization properties of the lowest Goldstone and Higgs
modes in the bulk phases are in full agreement with the general symmetry analysis \cite{GurarieChalker02,GurarieChalker03}
and with explicit results for Goldstone modes in a number of systems. As discussed at the end of Sec.\
\ref{sec:bogoliubov}, the fluctuation Hamiltonians have a structure equivalent to the chiral, orthogonal
symmetry class with $\nu=0$ the special reference energy. The extended character of the
zero-energy Goldstone mode agrees with the fact that the localization length diverges at the reference
energy for a chiral orthogonal random matrix ensemble. (This also implies that the correlations 
in the matrix elements of the
fluctuation Hamiltonians do not change the localization properties qualitatively.)

%%%%%%%%%%%%%%%%%%%%%%%%%%%%%%%%%%%%%%%%%%%%%%%%%%%%%%%%%%%%%%%%%%%%%%%%%%%%%%%%%%%%%
\subsection{Localization properties of higher Goldstone and Higgs excitations }
\label{subsec:higher_excitations_results}
%%%%%%%%%%%%%%%%%%%%%%%%%%%%%%%%%%%%%%%%%%%%%%%%%%%%%%%%%%%%%%%%%%%%%%%%%%%%%%%%%%%%%

We now turn to higher excitations in both the Goldstone and the Higgs channels.
In the Mott-insulating phase their behavior is easily understood. As all local order
parameters $\psi_j$ vanish in the Mott insulating phase, the disorder in the fluctuation
Hamiltonians $\mathcal{H}_G$ and $\mathcal{H}_H$ or, equivalently, the disorder
in the coupling matrices $\mathbf{X}_{G}$ and $\mathbf{X}_{H}$ is produced by the values of
 $U_i$ and $\tilde J_{ij}$ only.
The disorder is thus uncorrelated in space guaranteeing that all excitations in
the Mott-insulating phase are localized in two space dimensions.

The situation in the superfluid phase is more complicated because the coefficients of
$\mathcal{H}_G$ and $\mathcal{H}_H$ (or the matrix elements $X_{G,jk}$ and $X_{H,jk}$)
depend on the local mixing angles $\theta_j$. These angles are correlated because they
fulfill the mean-field equations (\ref{eq:sf_groundstate_condition}). Close to the
superfluid-Mott insulator transition, the correlations become long-ranged.
For correlated disorder, both extended and localized states are possible even in two
space dimensions.

\begin{figure}[t]
\includegraphics[width=6.5cm]{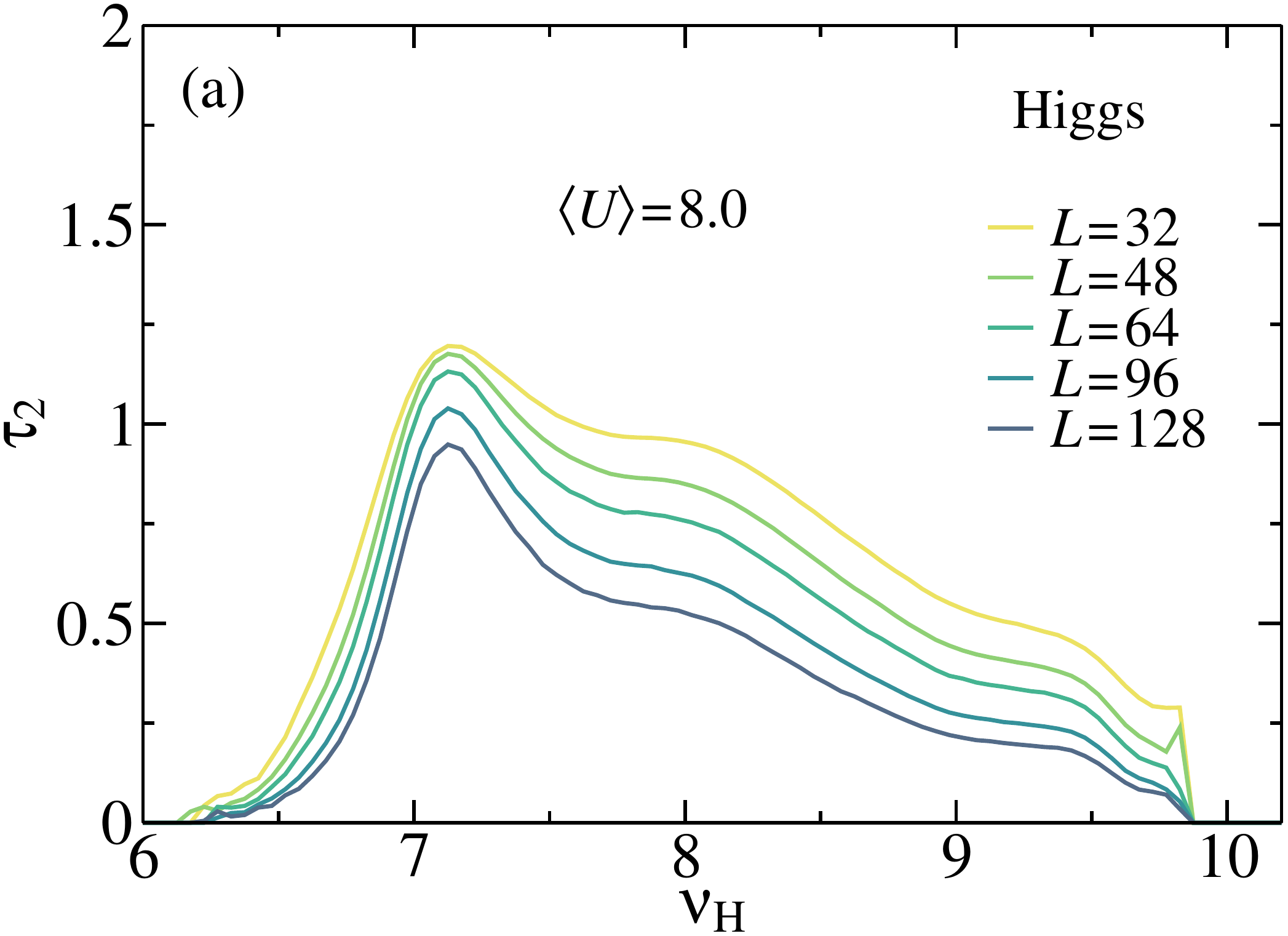}~
\includegraphics[width=6.5cm]{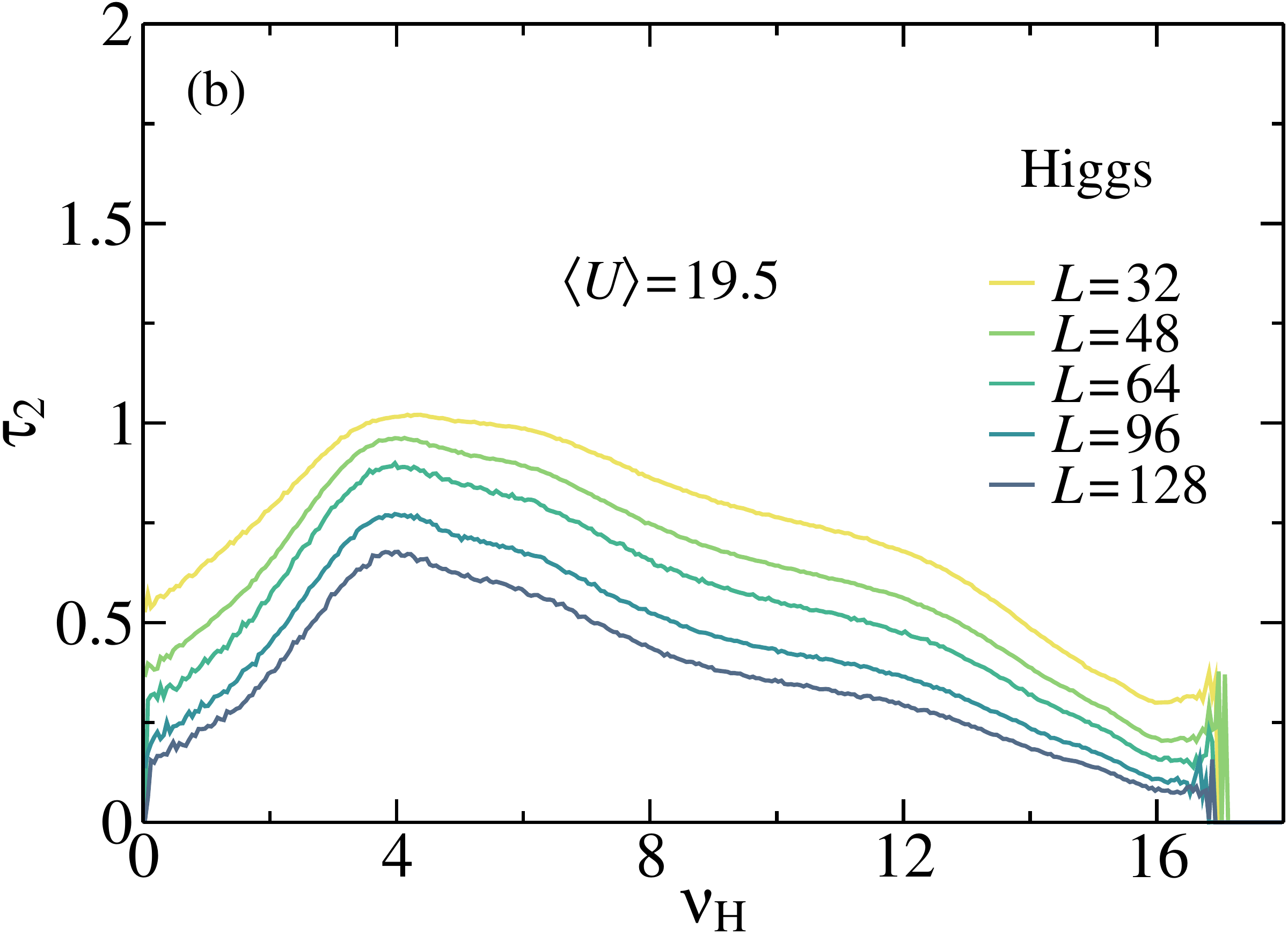}
\caption{Generalized dimension $\tau_2$ vs.\ excitation energy $\nu_H$ of the Higgs excitations for
two different values of $\langle U \rangle$ and several system sizes $L$. (Random-$U$ disorder of strength $r=1.0$.)
The data are averages over 1000 disorder realizations. Statistical errors are similar to the line widths. }
\label{fig:tau2_E_H_randomU}
\end{figure}

\begin{figure}
\includegraphics[width=8.5cm]{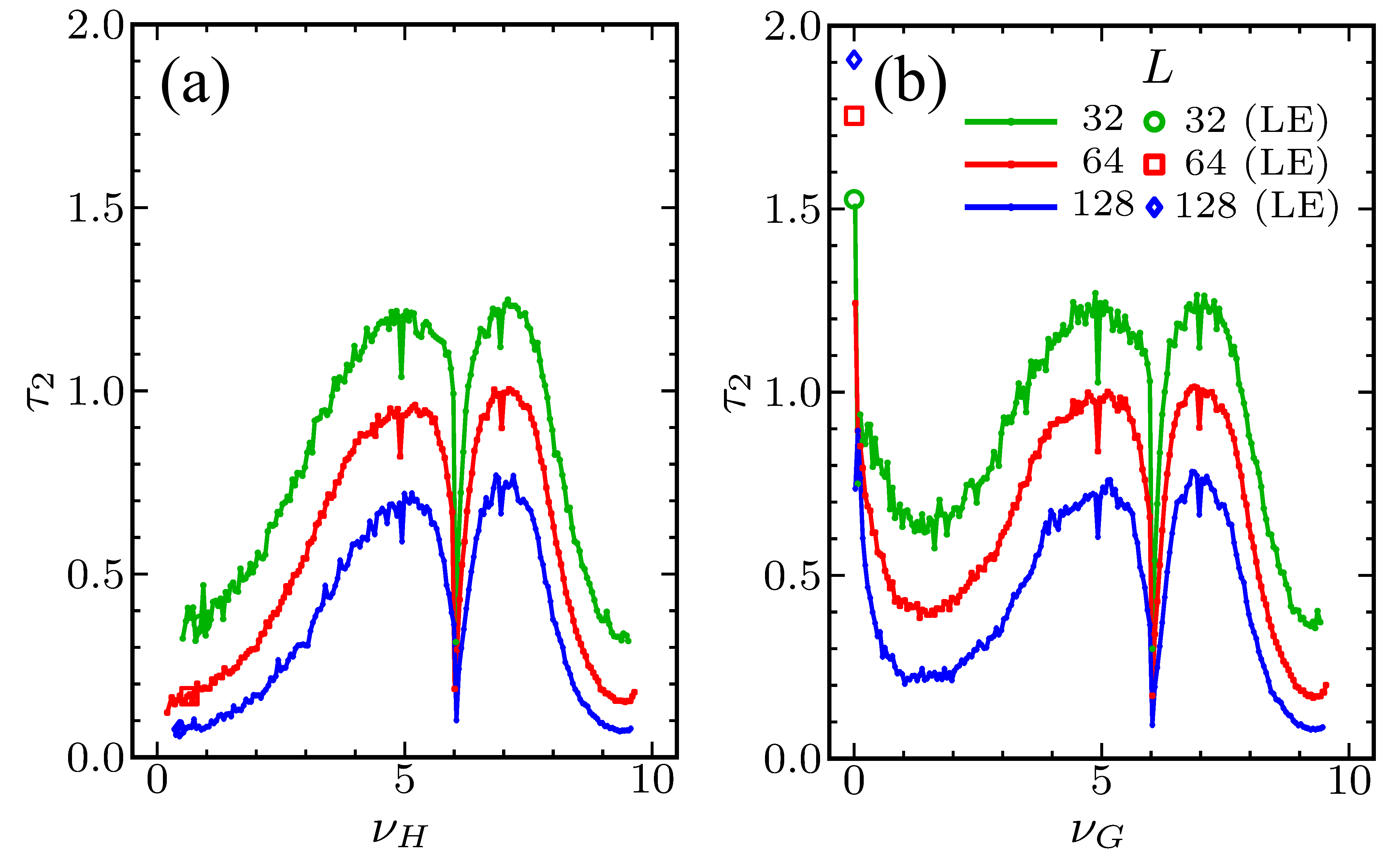}~
\includegraphics[width=4.2cm]{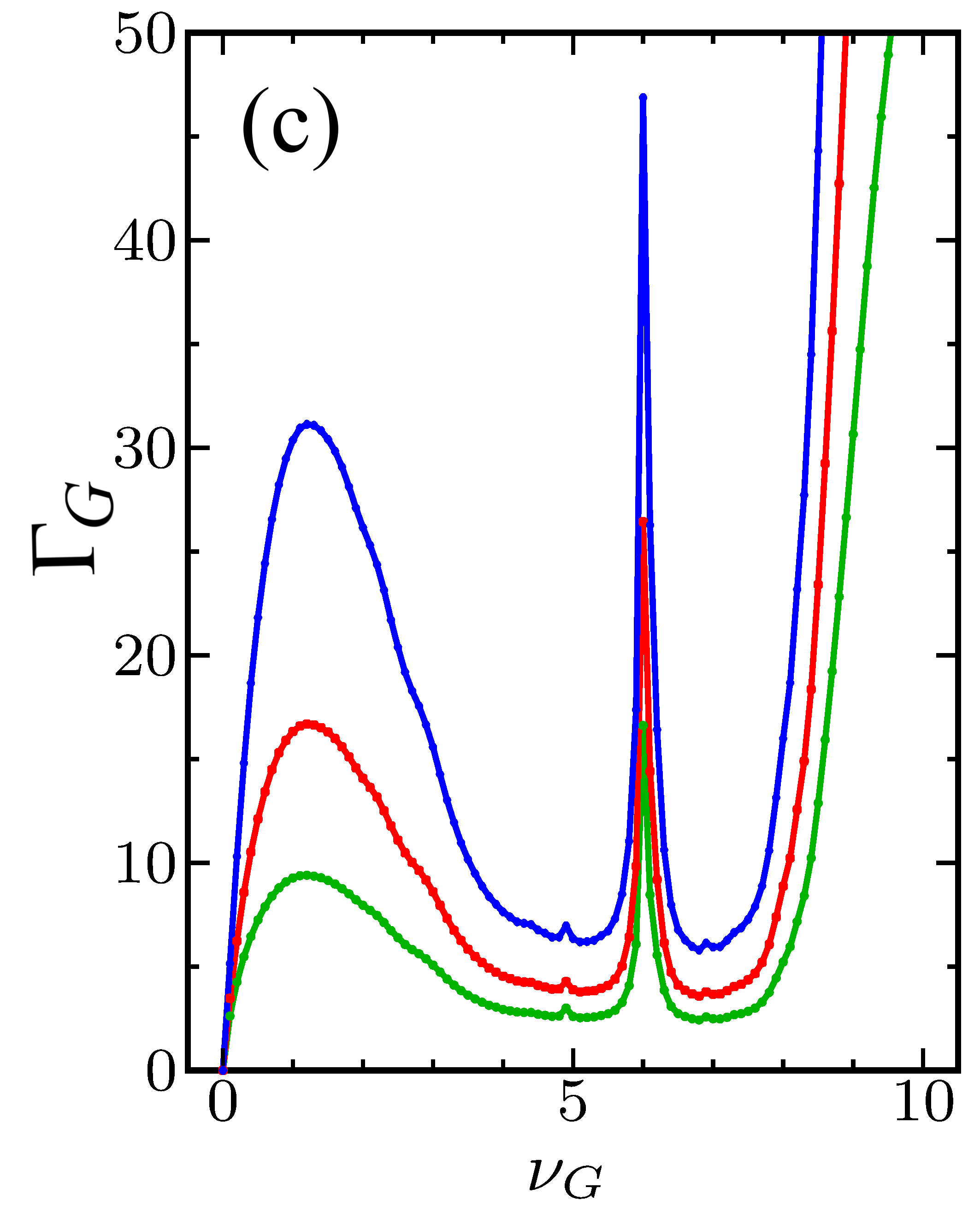}
\caption{Generalized dimension $\tau_2$ of Higgs (a) and Goldstone (b) excitations vs.\ excitation energy $\nu$
for $U=12$, dilution $p=1/3$ and several system sizes $L$ with $L/l=8$. The solid lines represent averages of
$\tau_2$ over small energy windows (width 0.1) and 100 to 400 disorder configurations, depending on $L$.
The values of $\tau_2$ of the lowest-energy
excitation (averaged over all disorder configurations) are shown as open symbols.
(c) Dimensionless Lyapunov exponent $\Gamma_G$ of the Goldstone excitations vs.\ excitation energy
$\nu_G$, calculated using the iterative Green's function method on strips of $L \times 10^6$ sites (the data
are averages over 12 strips)}
\label{fig:tau2_E_diluted}
\end{figure}

We first analyze the Higgs excitations in the superfluid phase. Figure \ref{fig:tau2_E_H_randomU}
presents the generalized dimension $\tau_2$ as a function of excitation energy $\nu_H$
for a system with random-$U$ disorder of strength $r=1$.
For both interaction strengths shown, viz., $\langle U \rangle = 8$ (deep in the superfluid phase)
and $\langle U \rangle = 19.5$ (close to the superfluid-Mott insulator transition),
$\tau_2$ decreases with increasing system size at all energies $\nu_H$. This implies that the
Higgs excitations are localized for all energies. We have obtained the same result for the
other studied strengths, $r=0.5$ and 1.5, of the random-$U$ disorder as wells as for dilution
disorder, as illustrated in Fig.\ \ref{fig:tau2_E_diluted}(a).

The Goldstone excitations in the superfluid phase display a more complex behavior. Figure
\ref{fig:tau2_E_diluted}(b) presents the generalized dimension $\tau_2$ of the Goldstone excitation
as a function of excitation energy $\nu_G$ for a diluted system with $p=1/3$ at $U=12$, slightly
inside the superfluid phase. The figure shows that $\tau_2$ for the lowest Goldstone excitation
increases with system size. This indicates an extended state in agreement with the results discussed
in Sec.\ \ref{subsec:lowest_excitations_results}. For all other excitation energies, $\tau_2$ decreases
with system size, implying that all Goldstone excitations except the lowest one are localized.
The same information can also be gained from Fig.\ \ref{fig:tau2_E_diluted}(c) which shows the
dimensionless Lyapunov exponent $\Gamma$ (calculated via the recursive Green function approach)
as a function of energy $\nu_G$.  $\Gamma$  increases with increasing strip width for all nonzero energies,
indicating that the Goldstone mode is localized. However, $\Gamma$ decreases rapidly
as the energy $\nu_G$ approaches zero, and for $\nu_G=0$, the Lyapunov exponent vanishes for all
strip widths, indicating an extended state.

\begin{figure}
\includegraphics[width=6.5cm]{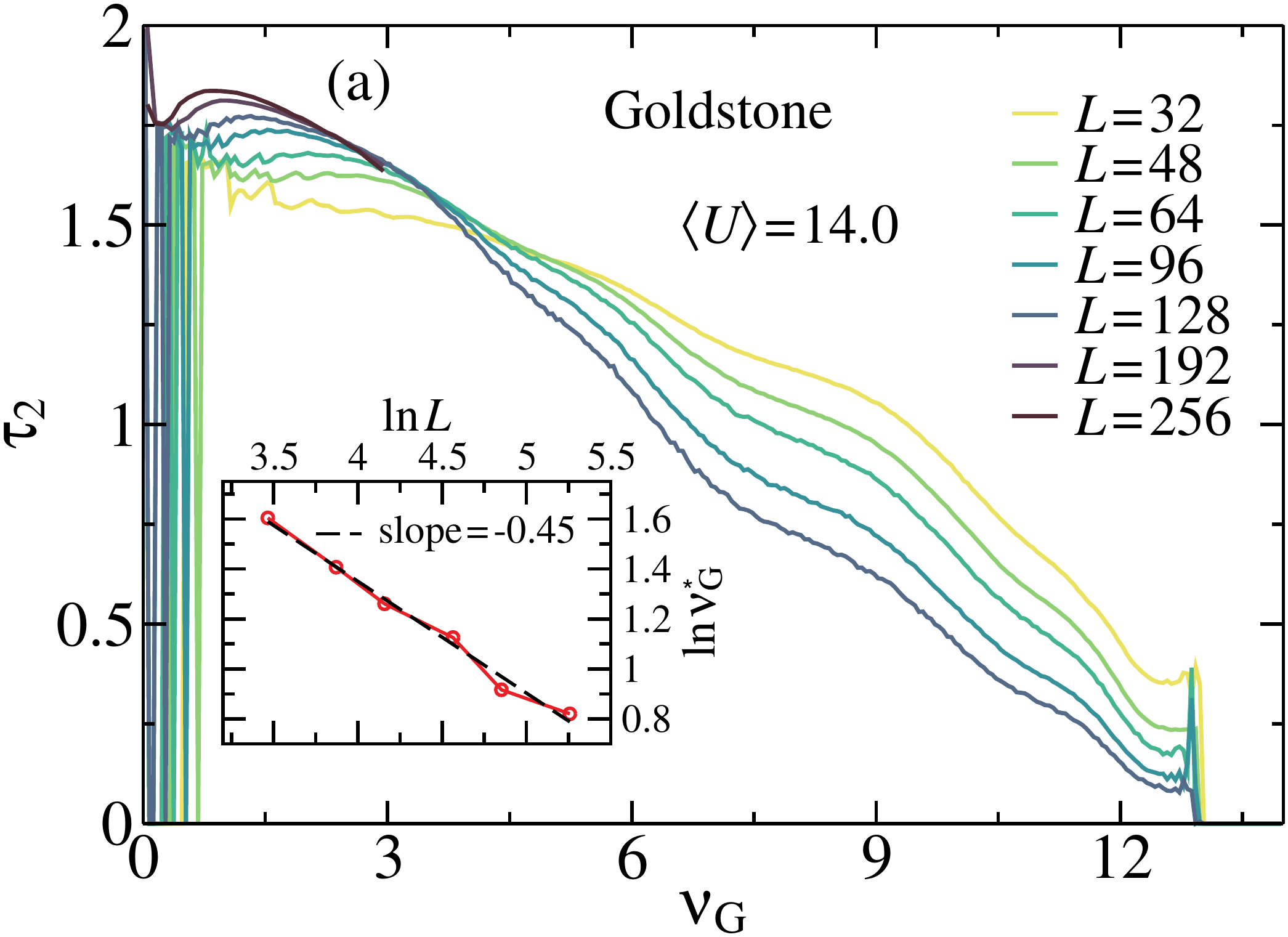}~\includegraphics[width=6.5cm]{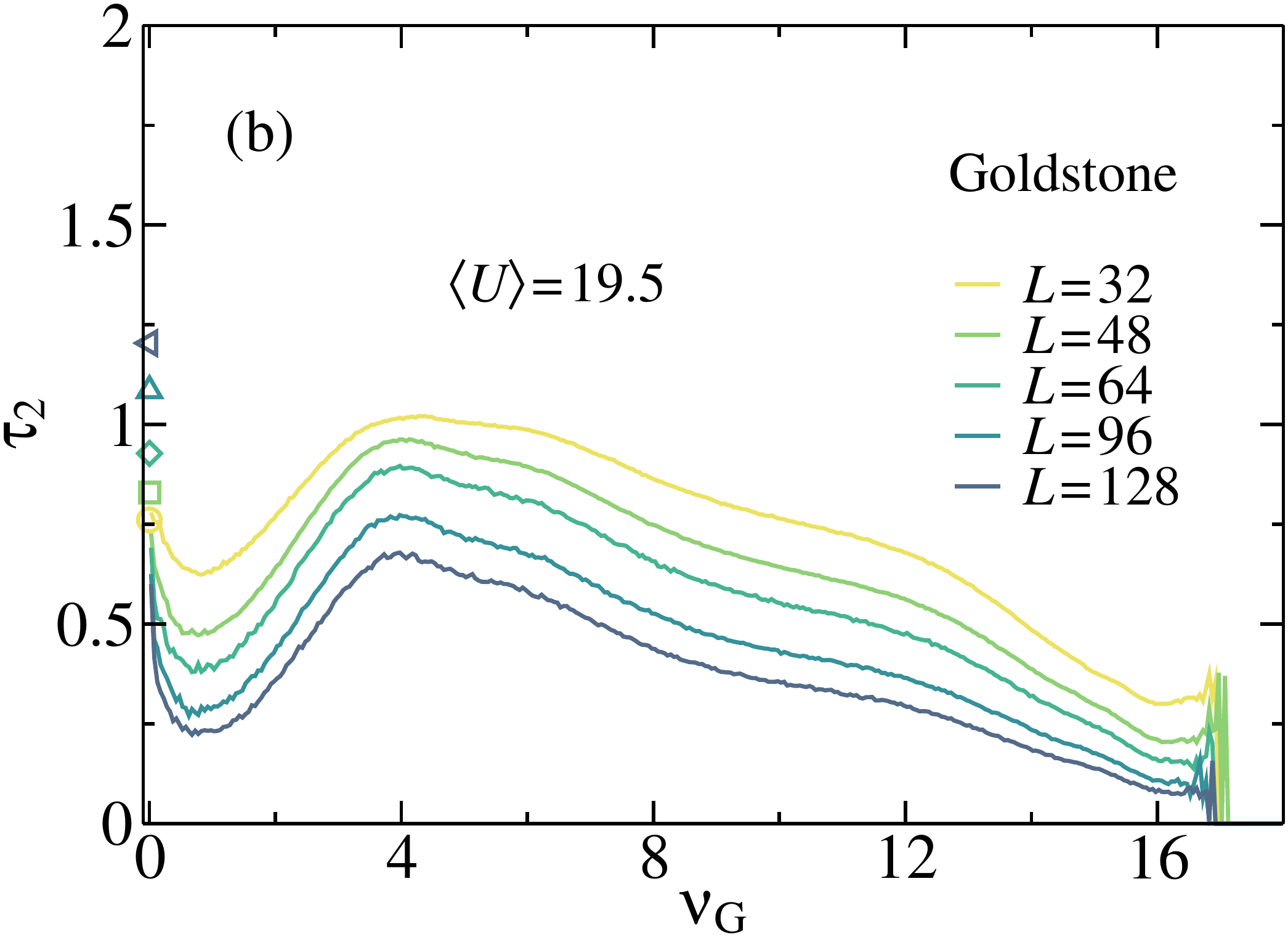}~
\caption{Generalized dimension $\tau_2$ vs.\ excitation energy $\nu_G$ of the Goldstone excitations for
several linear system sizes $L$ and two different values of $U$. (Random-$U$ disorder of strengths $r=1.0$.)
The data are averages over 1000 disorder realizations. Statistical errors are comparable to the line widths.
Inset: Crossing energy $\nu_G^*$ vs.\ linear system size $L$. }
\label{fig:tau2_E_G_randomU}
\end{figure}

We observe analogous behavior for other dilution values as well as for random-$U$ disorder, as illustrated
in Fig.\ \ref{fig:tau2_E_G_randomU}.
If the disorder is weak and/or the system is deep in the superfluid phase
[as in Fig.\ \ref{fig:tau2_E_G_randomU}(a)], the data seem to suggest -- at the first glance --
an entire range of energies with extended states because the $\tau_2$ vs.\ $\nu_G$ curves
for different system sizes display a crossing at a nonzero energy.
Below the crossing $\tau_2$ increases with size.
However, a more careful analysis shows that the crossing energy $\nu_G^*$ between the $\tau_2$ curves for two
consecutive sizes shifts towards $\nu_G=0$ with increasing size, as illustrated in the inset of Fig.\
\ref{fig:tau2_E_G_randomU}(a). In all such cases we have found no indication of the crossing points
converging to a nonzero energy.
We thus conclude that all Goldstone modes with nonzero excitation energy are localized in the
thermodynamic limit, even if the system is weakly disordered and/or deep inside the superfluid phase.
This is analogous to the behavior observed, e.g., for disordered phonons \cite{JohnSompolinskyStephen83}
who found the phonon localization length in a disordered elastic medium diverges as $\exp(1/\nu^2)$
in the low-frequency limit $\nu \to 0$.

It is interesting to compare Figs.\  \ref{fig:tau2_E_diluted}(a) and (b).
The Higgs and Goldstone modes show almost identical $\tau_2$ for larger excitation
energies, $\nu \gtrsim 3$, reflecting the fact that the two modes are still almost degenerate close
to the quantum phase transition. Analogously, the Higgs mode shown in Fig.\
\ref{fig:tau2_E_H_randomU}(b) and the Goldstone mode shown in Fig.\ \ref{fig:tau2_E_G_randomU}(b)
for the case of random-$U$ disorder feature the same $\tau_2$ for energies $\nu \gtrsim 4$.
Also note that the sharp features at energies around
$\nu=6$ observed in all panels of Fig.\ \ref{fig:tau2_E_diluted} are the result of the discrete
character of the site dilution disorder. They are absent for random-$U$ disorder (where the local
interactions are drawn from a continuous distribution).

%some heat maps of examples of exited states, if available

%%%%%%%%%%%%%%%%%%%%%%%%%%%%%%%%%%%%%%%%%%%%%%%%%%%%%%%%%%%%%%%%%%%%%%%%%%%%%%%%%%%%%
\subsection{Dynamic susceptibilities}
\label{subsec:susceptibilities_results}
%%%%%%%%%%%%%%%%%%%%%%%%%%%%%%%%%%%%%%%%%%%%%%%%%%%%%%%%%%%%%%%%%%%%%%%%%%%%%%%%%%%%%

\begin{figure}
\includegraphics[width=\textwidth]{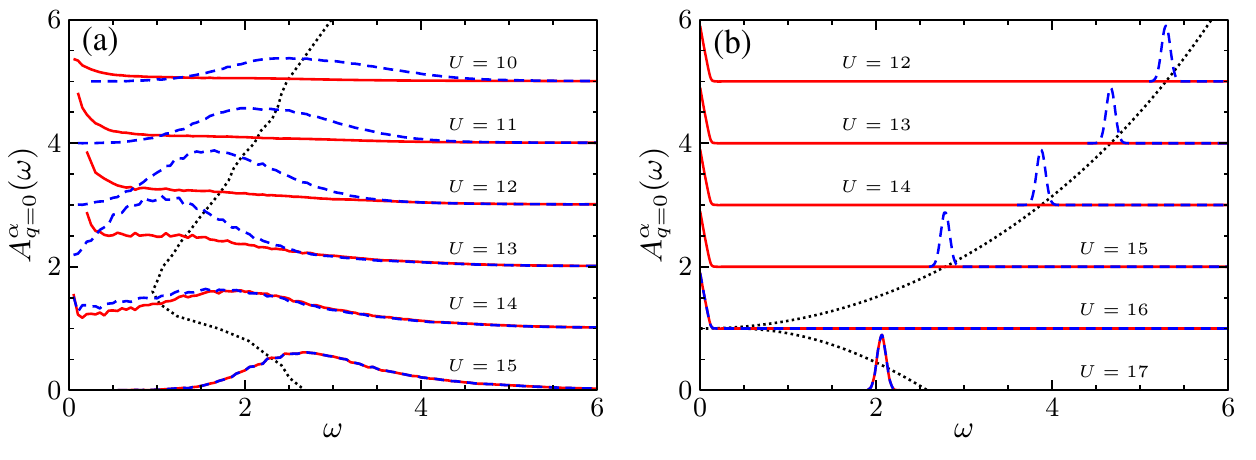}
\caption{Spectral functions $A^G_{\mathbf{q} = 0}(\omega)$ and $A^H_{\mathbf{q} = 0}(\omega)$
of the Goldstone (solid lines) Higgs (dashed lines) excitations, respectively,
for several interaction strengths $U$. The curves are shifted upwards with increasing $U$. Dotted lines mark the
position of the Higgs peak in $A^H$. (a) Dilution $p=1/3$ ($L=128$, 240 disorder realizations, statistical errors
are comparable to the line widths). (b) Clean case, $p=0$; here the peaks in the figure represent $\delta$ functions.
}
\label{fig:spectral_functions}
\end{figure}

To make direct contact with the Monte Carlo simulations of Ref.\ \cite{PuschmannCrewseHoyosVojta20},
we now analyze the spectral functions $A^G$ and $A^H$ associated with the Goldstone and Higgs
Green functions (\ref{eq:Green_G}) and (\ref{eq:Green_H}), respectively.
Figure \ref{fig:spectral_functions} presents these spectral functions, computed from the
mean-field eigenenergies and eigenstates using (\ref{eq:A_q}),
at zero wave vector for several interactions $U$, contrasting the clean case with
dilution $p=1/3$.
The spectral functions of the diluted system are very broad, even though the (single-particle) eigenstates
of $\mathcal{H}_G$ and $\mathcal{H}_H$ are
noninteracting within the Gaussian approximation and thus have no intrinsic width. This broadening
stems from the fact that many localized eigenstates contribute to the zero-wave-vector spectral function,
i.e., it is caused by disorder-induced localization effects. We also observe that the peak energy $\omega_H$
of the Higgs spectral function does not soften at the superfluid-Mott insulator transition, in contrast
to what is expected from naive scaling. This mirrors the Monte Carlo results.
(Note that the peak energy $\omega_H$ differs from the energy gap $m_H$ that marks the lowest excitation energy.)
In contrast, the clean spectral functions show the expected $\delta$ peaks at energies corresponding
to the Higgs and Goldstone masses.

\begin{figure}
\includegraphics[width=6.5cm]{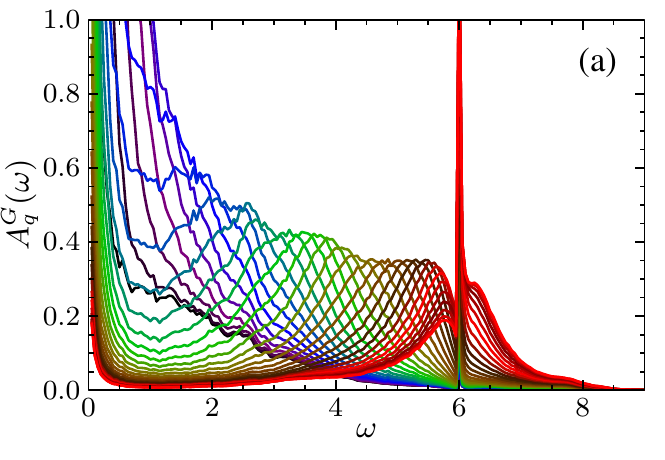}~
\includegraphics[width=6.5cm]{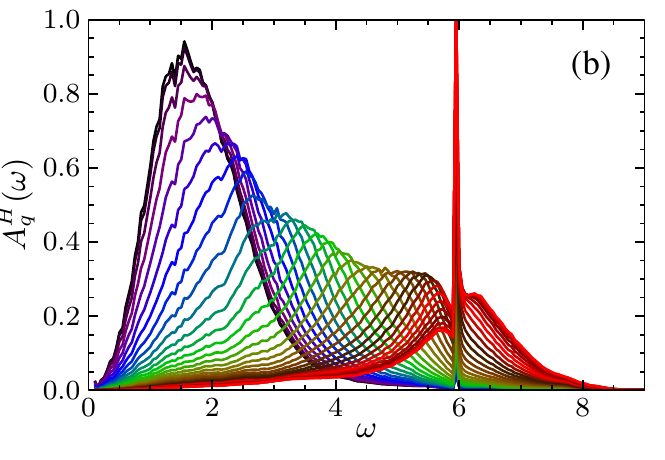}
\caption{Spectral functions $A^G_\mathbf{q}(\omega)$ and $A^H_\mathbf{q}(\omega)$ vs.\ $\omega$
for a diluted system with $p=1/3$ and $U=12$  ($L=128$, 240 disorder realizations, statistical errors
are comparable to the line widths). The wavevector $q_x$ varies from $q_x=0$ (black)
to $q_x=\pi$ (red) in 32 steps while $q_y$ is fixed at zero.}
\label{fig:spectral_functions_q}
\end{figure}

\begin{figure}
\includegraphics[width=6.5cm]{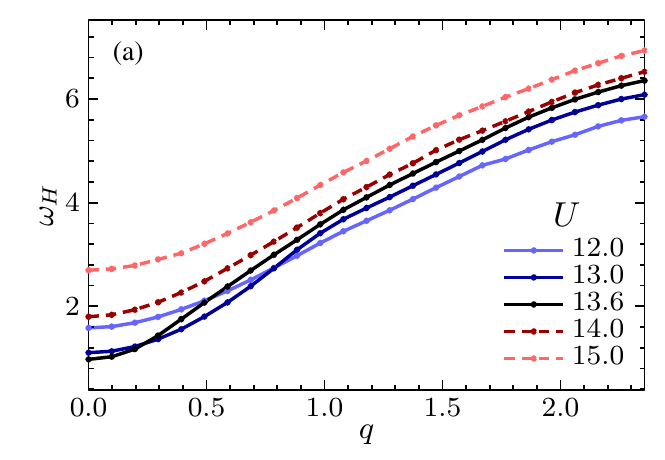}~
\includegraphics[width=6.5cm]{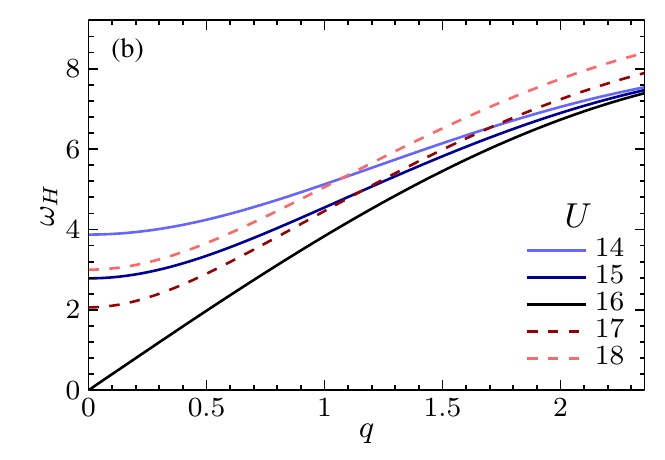}
\caption{(a) Peak position $\omega_H$ of the Higgs spectral function $A^H_\mathbf{q}(\omega)$
shown in Fig.\ \ref{fig:spectral_functions_q} vs.\ wave vector $q$ (along the coordinate directions).
(b) $\omega_H$ vs.\ $q$ for the clean case.
}
\label{fig:Higgs_dispersion}
\end{figure}

We also investigate the $q$-dependence of the Goldstone and Higgs spectral functions. Figure
\ref{fig:spectral_functions_q} presents $A^G_\mathbf{q}(\omega)$ and $A^H_\mathbf{q}(\omega)$
for a diluted system ($p=1/3$) slightly inside the superfluid phase.
The Higgs spectral function $A^H_\mathbf{q}(\omega)$ further broadens with increasing $q$, and the peak energy $\omega_H$
increases. The resulting dispersion relation $\omega_H(q)$ is presented in Fig.\ \ref{fig:Higgs_dispersion}
and compared to that of the clean system.
In the clean case, the data show the behavior expected for a $z=1$ quantum critical point. The low-energy dispersion
is linear, $\omega_H \sim |\mathbf q|$, at criticality $U=U_{c0}=16$. Away from criticality,
it crosses over to the quadratic form
$\omega_H(q) = \omega_H(0) +c \mathbf q^2$. In contrast, the dispersion of the diluted system does not change much with
the distance from criticality, and features always a quadratic $q$-dependence for small $q$.
The Goldstone spectral function $A^G_\mathbf{q}(\omega)$ also develops a finite-energy peak for larger $q$.
However, even at the largest $q$, $A^G_\mathbf{q}(\omega)$ retains a zero-energy peak, albeit of reduced amplitude.
This suggests that the lowest (zero-energy) Goldstone excitation has nonzero Fourier components in the
entire Brillouin zone.

%%%%%%%%%%%%%%%%%%%%%%%%%%%%%%%%%%%%%%%%%%%%%%%%%%%%%%%%%%%%%%%%%%%%%%%%%%%%%%%%%%%%%
\section{Conclusions}
\label{sec:conclusions}
%%%%%%%%%%%%%%%%%%%%%%%%%%%%%%%%%%%%%%%%%%%%%%%%%%%%%%%%%%%%%%%%%%%%%%%%%%%%%%%%%%%%%

In summary, we have developed a quantum mean-field theory for a model of disordered and
interacting bosons to complement recent quantum Monte Carlo simulations of the collective
modes near the superfluid-Mott glass quantum phase transition. The theory
describes the quantum ground state as an inhomogeneous
product wave function. The mean-field theory needs to be solved numerically,
but it is able to capture the spatial inhomogeneities of the superfluid order parameter.
Collective excitations are then obtained from an expansion of the Hamiltonian
in the fluctuations about the mean-field ground state up to quadratic order.
Extensive numerical calculations have demonstrated that all excitations are spatially
localized in the Mott-insulating phase (where the two modes are degenerate because the
$U(1)$ order parameter symmetry is not broken) for arbitrary excitation energy.
The Higgs mode is still localized for all energies in the superfluid phase.
In contrast, the lowest (zero-energy) Goldstone excitation delocalizes in the superfluid
phase whereas all higher-energy Goldstone excitations are localized. This implies that
there is no mobility edge for the excitations at a nonzero energy. These qualitative
localization properties agree with the general symmetry classification for localization
of bosonic excitations \cite{GurarieChalker03}.

We have also computed the spectral functions of the collective excitations. The Higgs spectral
function $A^H(\omega)$ displays a broad maximum whose peak frequency remains nonzero across
the transition, resembling the corresponding Monte Carlo results \cite{PuschmannCrewseHoyosVojta20}.
As the excitations are non-interacting within the mean-field theory, this provides evidence
for the unconventional (noncritical) scalar response being (mainly) caused by localization effects
rather than enhanced damping due to mode-mode coupling.

In this concluding section we put our results into a broader perspective, and we discuss
open questions.
Let us comment on the validity of the mean-field approach. Based on a comparison with
recent Monte Carlo simulations \cite{Vojtaetal16,PuschmannVojta17,LerchVojta19},
the qualitative thermodynamic properties of the model are well described by the mean-field solution
outside of the immediate vicinity of the superfluid-Mott insulator quantum phase transition.
The mean-field ground state in the superfluid phase features a spatially inhomogeneous
local order parameter. The relative spatial variations of the order parameter are small deep
inside the phase but they grow as the quantum phase transition is approached.
Unfortunately (but not unexpectedly), the mean-field approach fails to correctly describe
the Mott glass, i.e., the quantum Griffiths phase of the superfluid-Mott insulator transition
in which isolated superfluid puddles coexist with an insulating bulk. As the mean-field theory
cannot account for order parameter fluctuations, it assigns a static superfluid order parameter
to these puddles (rare regions), effectively replacing the quantum Griffiths phase with the
tail of a smeared quantum phase transition. If it were correct, such an exotic smeared quantum phase transition
would be an exciting finding. Here it is a mean-field artifact as the superfluid-insulator quantum phase transition in our
model has been shown to be conventional by the above-mentioned Monte Carlo simulations,
in agreement with the general classification of disordered quantum phase transitions
\cite{VojtaSchmalian05,Vojta06,VojtaHoyos14}.
One important consequence of this issue is that we are unable to locate the exact positions of both the
thermodynamic quantum phase transitions and the delocalization transition of the Goldstone mode.

Despite these limitations, some key results appear to be robust and independent of the mean-field
approximation. Importantly, the scalar response (Higgs spectral function) is dominated by a
broad peak at rather high (microscopic) frequencies that changes only little as the system is tuned from
insulator through the critical region into the superfluid. It it thus unlikely that it is significantly
affected by the intricacies of the critical behavior close to the transition.
Moreover, as pointed out above the mean-field results match the behavior of the numerically
exact Monte Carlo simulations of the excitations \cite{PuschmannCrewseHoyosVojta20}.

The quadratic (Gaussian) approximation of the fluctuation Hamiltonians $\mathcal{H}_G$ and
$\mathcal{H}_H$ does not take anharmonic (mode-coupling) effects into account. They could be included
by keeping higher-order terms in the expansion of the Hamiltonian and would allow us to study the
effects of damping. Exploring their influence
and the interplay between the anharmonicities and the disorder remains a task for the future.

As pointed out in Ref.\ \cite{GurarieChalker03}, one of the important differences between
fermionic and bosonic localization problems stems from the fact that thermodynamic stability
requires all bosonic excitation energies to be positive. This generally requires specific
correlations of the Hamiltonian matrix elements.
In our system, the disorder in the fluctuation Hamiltonians in the Mott insulating phase is
uncorrelated because all local order parameters $\psi_j$ vanish and the matrix elements depend
on the uncorrelated random variables $U_j$ and $\tilde J_{jk}$ only.
Thus, the localization of all excitations in the Mott insulating phase can be
understood as direct consequence of localization physics in the conventional orthogonal symmetry class
(or, equivalently in the chiral orthogonal class at energies away from the reference
energy $\nu=0$) \cite{EversMirlin08}.
In the superfluid phase, in contrast, the disorder in the fluctuation Hamiltonians
is correlated because it depends on the inhomogeneous local order parameters $\psi_j$
which are the solutions of the coupled mean-field equations (\ref{eq:sf_groundstate_condition})
and thus correlated. Deep inside the superfluid phase, the correlations between the  $\psi_j$
are short-ranged as they are governed by the thermodynamic correlation length. However,
this correlation length diverges as the quantum critical point is approached.
On the one hand, these correlations of the matrix elements preserve the correct bosonic character
of the spectrum related to the chiral symmetry discussed in Sec.\ \ref{sec:bogoliubov}
and lead to the delocalization of the zero-energy Goldstone excitation.
One the other hand, these correlations could in principle lead to nontrivial modifications
of the localization behavior compare to the corresponding chiral random matrix ensemble,
in particular at higher energies. Our numerical results do not show indications of such modifications.

Within the quadratic (Gaussian) approximation, the fluctuation Hamiltonians describe
non-interacting quasi particles. It is interesting to ask whether concepts of many-body
localization become important for the collective excitations once anharmonic terms
beyond the quadratic approximation are included. Specifically, will mode coupling
effects have delocalizing tendencies for higher excitations in the middle of the energy band
where the density of states is large and anharmonic terms should be able to couple
a large number of states?

In this paper, we have focused on the two-dimensional Bose-Hubbard Hamiltonian.
However, the mean-field approach can be applied to three-dimensional systems as well.
The thermodynamic critical behavior of the superfluid-Mott glass quantum phase
transition in three dimensions has recently been shown to be conventional and
similar to the two-dimensional transition \cite{CrewseLerchVojta18}. It will be interesting
to study the collective excitations in this case. Some work along these
lines is already in progress. In addition, the mean-field approach can be generalized to
other problems such as the generic superfluid-Bose glass transition.

In conclusion, our results show that disordered quantum phase transitions can feature unconventional
collective excitations that violate naive scaling even if their thermodynamic critical behavior is entirely conventional.
This raises the question of whether or not one can classify the excitation dynamics
of disordered quantum phase transitions along similar lines as their thermodynamics
\cite{VojtaSchmalian05,Vojta06,VojtaHoyos14}.

%%%%%%%%%%%%%%%%%%%%%%%%%%%%%%%%%%%%%%%%%%%%%%%%%%%%%%%%%%%%%%%%%%%%%%%%%%%%%%%%%%%%%
\section*{Acknowledgements}
%%%%%%%%%%%%%%%%%%%%%%%%%%%%%%%%%%%%%%%%%%%%%%%%%%%%%%%%%%%%%%%%%%%%%%%%%%%%%%%%%%%%%

This work was supported by the NSF under Grant Nos.\ DMR-1506152, DMR-1828489, PHY-1607611, and OAC-1919789 (T.V.),
by Conselho Nacional de Desenvolvimento Cient\'{\i}fico e Tecnol\'{o}gico (CNPq) under Grant No. 312352/2018-2, and by
FAPESP under Grant Nos. 2015/23849-7 and 2016/10826-1 (J.H.).
M.P.\ acknowledges support from the German Research Foundation (DFG) through the
Collaborative Research Center, Project ID 314695032 SFB 1277 (projects A03, B01).
T.V. and J.A.H. acknowledge the hospitality of the Aspen Center for Physics, and T.V. thanks the
Kavli Institute for Theoretical Physics where part of the work was performed.

\appendix
%%%%%%%%%%%%%%%%%%%%%%%%%%%%%%%%%%%%%%%%%%%%%%%%%%%%%%%%%%%%%%%%%%%%%%%%%%%%%%%%%%%%%
\section{Numerical solution of the mean-field equations}
\label{app:mf_numerics}
%%%%%%%%%%%%%%%%%%%%%%%%%%%%%%%%%%%%%%%%%%%%%%%%%%%%%%%%%%%%%%%%%%%%%%%%%%%%%%%%%%%%%

The first step in our approach is the numerical solution of the mean-field equations
(\ref{eq:sf_groundstate_condition}) which constitute a large system of coupled nonlinear
equations. We implement two different numerical algorithms to solve these equations
efficiently and accurately, a simple iteration and a gradient descent method.

The first method consists of rewriting the mean-field equations (\ref{eq:sf_groundstate_condition})
in the form
\begin{equation}
    \S{i} =	(4/U_i) \sqrt{1-\Ssq{i}} \sum_{j} \tilde J_{ij} \S{j}  ~.
\label{eq:gs_iteration}
\end{equation}
It then starts from a guess for the local order parameters $\psi_i = \S{i}$, for example random numbers
between 0 and 1. Inserting these values on the r.h.s.\ of (\ref{eq:gs_iteration})
produces new values on the l.h.s.\ of the equation. This step is iterated until
the difference between the old and new values falls below an accuracy threshold.
As our model does not contain competing interactions, the mean-field ground state is unique
(up to the overall phase which we have fixed at zero). The iterative method reliably
converges to the ground state even though the convergence becomes very slow close to the
superfluid-Mott insulator transition.

The gradient descent method numerically minimizes the mean-field ground state energy
(\ref{eq:thermodnamic_groundstate}),
\begin{equation}
E_0 = -\frac{1}{2} \sum_{ij} \tilde{J}_{ij} \S{i} \S{j} + \frac{1}{2} \sum_i U_i \Shsq{i},
\end{equation}
with respect to local order parameters $ \S{i}$. In each step of the method, we obtain an improved set of order
parameters by finding the local energy minimum along the gradient direction $-\partial E_0/\partial  \S{i}$. This approach converges when no energy reduction can be achieved within a prescribed accuracy.

Interestingly, the simple iterative solution of (\ref{eq:gs_iteration}) turns out to be more accurate, in particular in situations where the local order parameters are close to zero. The gradient method loses the ability to discriminate
between states that are close in energy, mainly due to the numerical errors in the
transformation between $\S{i}$ and $\C{i}$ and the computation of energy differences.
The strongest impact of the numerical inaccuracies occurs in the Griffith region, where superfluid puddles
in a insulating matrix lead to differences of several orders of magnitudes in the local order parameter.
In the Griffiths region, the energy difference between the lowest and second-lowest Goldstone excitations
can become extremely small, and the numerical inaccuracies can cause them to switch places.
If that happens, the seeming (but incorrect) lowest Goldstone excitation disagrees with the analytical
expression (\ref{eq:HandySolution}).
As the two lowest Goldstone excitations can have very different localization properties, such
switches introduce sizable errors in the results.
Using quadruple precision real variables (16 byte) instead of double precision (8 bytes) alleviates this
problem at the price
of significantly slower performance. Moreover, the exact expression (\ref{eq:HandySolution})
for the ground state can be used to check the numerics. The iteration method, fulfilling the mean-field
equations locally with higher accuracy, does not suffer from this effect but is very slow in the
Griffiths phase.

To overcome these difficulties, we often compute the mean-field ground state using quadruple precision
but perform the consecutive analysis of the excitations in double precision. In this way, the
coupling matrices $X_{G,ij}$ and $X_{H,ij}$ in the eigenvalue problem for the excitations are effectively
free of numerical errors (within double precision) which turns out to be sufficient for reliable results.

%%%%%%%%%%%%%%%%%%%%%%%%%%%%%%%%%%%%%%%%%%%%%%%%%%%%%%%%%%%%%%%%%%%%%%%%%%%%%%%%%%%%%
\section{Local order parameters and eigenstates for the case of random $U$ disorder}
\label{app:randomU}
%%%%%%%%%%%%%%%%%%%%%%%%%%%%%%%%%%%%%%%%%%%%%%%%%%%%%%%%%%%%%%%%%%%%%%%%%%%%%%%%%%%%%

\begin{figure}
\includegraphics[width=\textwidth]{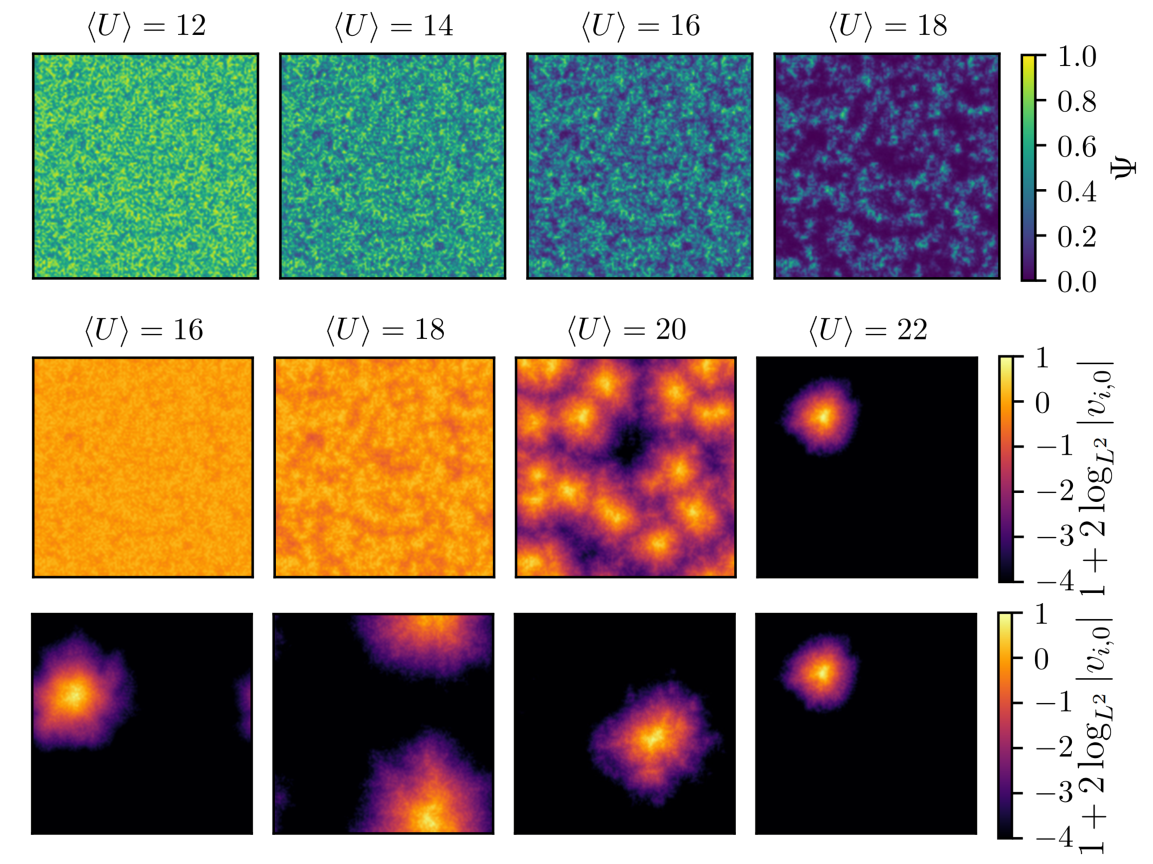}
\caption{Local order parameters $\psi_j$ (top row), and eigenstate wavefunctions
$\mathcal{V}_{Gj0}$ and $\mathcal{V}_{Hj0}$ of the lowest Goldstone (middle row)
and Higgs (bottom row) excitations of a single
disorder realization ($L=128$, random-$U$ disorder with $r=1.0$, and different $U$).}
\label{fig:LocalOPandGS_randomU}
\end{figure}

Figure \ref{fig:LocalOPandGS_randomU} presents the local order parameters $\psi_j$ and the wave functions of the
lowest Goldstone and Higgs excitations for a single disorder realization of a system with random-$U$ disorder ($r=1$).
The qualitative features are similar to those of the diluted system shown in Fig.\ \ref{fig:LocalOPandGS}.
The order parameter features superfluid puddles in an insulating bulk close to the onset of superfluidity.
Deeper inside  the superfluid phase , the order parameter is only weakly inhomogeneous.
The lowest Higgs excitation is strongly localized for all $U$ whereas the lowest Goldstone excitation
undergoes a delocalization transition when the system enters the superfluid phase.

\bibliographystyle{elsarticle-num-names}
\bibliography{../00BibTeX/rareregions}

\end{document}